\documentclass[amsfonts,aps,nofootinbib,preprint,superscriptaddress]{revtex4}
\usepackage{epsf}  
\usepackage{epsfig}
\usepackage{graphicx}

\begin{document}

\title{Supergraph Approach in a Higher-order LDE Calculation of the Effective Potential for F-type Broken SUSY}

\author{M. C. B. Abdalla{\footnote{mabdalla@ift.unesp.br}}}
\affiliation{Instituto de F\'{\i}sica Te\'{o}rica, UNESP - Universidade Estadual Paulista, Rua Dr. Bento Teobaldo Ferraz 271, 
Bloco II, Barra-Funda, Caixa Postal 70532-2, 01156-970, S\~{a}o Paulo, SP, Brazil}
\author{R. L. S. Farias{\footnote{ricardofarias@ufsj.edu.br}}}
\affiliation{Departamento de Ci\^encias Naturais, Universidade Federal de S\~ao Jo\~ao del Rei, 36301-000, S\~ao Jo\~ao del Rei, MG, Brazil}
\author{J. A. Helay\"{e}l-Neto{\footnote{helayel@cbpf.br}}}
\affiliation{Centro Brasileiro de Pesquisas F\'isicas, Rua Dr. Xavier Sigaud 150, 
Urca, Rio de Janeiro, RJ, 22290-180, Brazil}
\author{Daniel L. Nedel{\footnote{daniel.nedel@unipampa.edu.br}}}
\affiliation{Universidade Federal do Pampa, 
Rua Carlos Barbosa S/N, Bairro Get\'ulio Vargas, 96412-420, Bag\'e, RS, Brazil}
\author{Carlos R. Senise Jr.{\footnote{carlossenise@unipampa.edu.br}}}
\affiliation{Universidade Federal do Pampa, Av. Pedro Anuncia\c{c}\~ao S/N, Vila Batista, 96570-000, Ca\c{c}apava do Sul, RS, Brazil}

\begin{abstract}

In this work, we adopt the simplest model that spontaneously breaks supersymmetry, namely, 
the minimal O'Raifeartaigh model. The effective potential is computed in the framework of the 
linear delta expansion (LDE) approach up to the order $\delta^2$, conjugated with superspace and 
supergraph techniques. The latter can be duly mastered even if supersymmetry is no longer exact 
and the efficacy of the superfield approach in connection with the LDE procedure is confirmed 
according to our investigation. That opens up a way for a semi-nonperturbative superspace 
computation which allows us to deal with spontaneously broken supersymmetric models and 
encourages us to go further and apply this treatment to the Minimal Supersymmetric Standard 
Model (MSSM) precision tests.
\end{abstract}

\maketitle


\section{Introduction}

The thrilling times of the LHC Physics we are living in opens up a great deal of issues connected to fundamental mechanisms and theories, in special, supersymmetry (SUSY) and its possible breaking mechanisms~\cite{O'Raifeartaigh,Fayet-Iliopoulos,NS,IS,ISS1,ISS2,Shih,Marques,delta1,Nibbelink}. Considering fundamental principles of quantum fields theory, SUSY seems to be fairly well motivated as a very fundamental symmetry of the high-energy regime. At our accessible energies, it does not show up; it has to be broken at some scale much above our reachable energies, and its possible evidences at accelerator energies must be communicated by means of some mechanism connecting this (higher energy) breaking scale to our low-energy world.

The present and the near-future LHC outcomes are crucial for the interplay between SUSY and the Standard Model parameters. The focus is not on SUSY itself, once we understand that SUSY is very likely to show up at very high-energy scales; the actual matter with SUSY relies on its possible breaking mechanisms and the ways the latter are driven down to the cutoff region of the Standard Model, namely, the TeV scale. In this framework, the quest for possible new SUSY violation mechanisms and a broader exploitation of the already known models to breakdown the fermion/boson symmetry are self justifiable \cite{susybreaking}.

On the other hand, LHC Physics is also refining the precision tests and the level of accuracy of the measurements of the Standard Model's parameters. Since perturbative quantum field-theoretic calculations are the way we get the phenomenological results for Standard Model processes to be compared with experiments, we are face to face with the need to go further in perturbation theory, so as to incorporate higher order corrections into the calculation of physical processes.

Placed in this scenario, we are motivated to reassess SUSY breaking models by computing higher order corrections to their corresponding effective potentials, so as to probe the effects of SUSY breaking in connection with the improvement of precision tests at the LHC. It is clear that LHC is a collider for new discoveries rather than a precision machine; but, anyhow, it increases the level of the precision tests of the previous LEP. It becomes a mandatory task to ascertain how much loop corrections affect the pattern of SUSY breaking once we start  off from a violation that takes place at the classical level.

In connection with the discussion of SUSY breaking to account for the splitting of the masses of supersymmetric partners, we would like to point out that, very recently, a new proposal of a model based on SUSY has been proposed by Alvarez, Valenzuela and Zanelli \cite{alvarez}, in which a structure of partners do not show up,  although SUSY is locally realised. So, there is no need of SUSY breaking to split masses and the (fermionic) matter fields acquire mass through their coupling with some background geometry. 

In a series of previous works \cite{OFTdelta,renormOR}, we have adopted the minimal O'Raifeartaigh model \cite{O'Raifeartaigh}, which realises SUSY violation by means of the so-called F-terms, and we have deviced a technique to approach the problem with the use of superfield and supergraph techniques. To get a richer perturbative series, we have chosen the so-called linear delta expansion (LDE)\footnote{in recent studies in the literature some authors have called the LDE method as optimized perturbation theory (OPT), since the method is not only an expansion in the parameter $\delta$, there is an optimization procedure on the method.} and we have coupled this method to our superfield methods. The outcome was encouraging and, in view of the efficacy of the conjugating supergraph techniques with the LDE, here we propose to carry out a computation of the effective potential up to $\mathcal{O}(\delta^2)$. Owing to a particularity of the LDE, calculations at this order require to take into account vacuum diagrams up to two loops.

The LDE is a nonperturbative method which automatically resums large classes of terms in a self-consistent way, to avoid possible dangerous overcounting of diagrams. This is achieved by combining perturbation theory with an optimization procedure. It has a long history of successful applications, describing phenomenological models using quantum field theory at zero temperature and under extreme conditions. It has been shown that the LDE results go beyond the standard mean-field or large-$N$ approximation by explicitly including finite-$N$ effects. Some very interesting results can be founded in~\cite{krein,pintoramosphi4,tricritical,caldas,PRD2008farias,optbtphi4,NJLSU2,OPT,bulkOPT} and in references therein, and strong signals of the convergence of the method can be found in \cite{braaten,prlbec}.

Usually, when two (or more) loops are present in the perturbation series, we need to implement a numerical calculation to perform the optimization procedure. In this case, it was shown in all applications cited above that the numerical results of the LDE can go beyond the usual resumation methods. Here we further develop the superspace applications of the LDE, by taking into account $\mathcal{O}(\delta^2)$ terms in the effective potential expansion and solving numerically the optimization procedure. In particular we study the convergence of the method, where we contrast the numerical results with our analytical results obtained at $\mathcal{O}(\delta^1)$ using two different optimization procedures.

The general structure of our paper is as follows: in Section II, we briefly review the application of the LDE to supersymmetric field theories, while working in superspace. In Section III, we employ supergraph techniques to compute the one- and two-loop diagrams that contribute to the $\mathcal{O}(\delta^2)$ to the effective potential. All the perturbative calculations of Section III and the numerical results we work out are collected in Section IV. Finally, our Concluding Remarks are cast in Section V. An Appendix follows, where we list all the superspace integrals of the supergraphs evaluated in Section III.


\section{Catching-up of Superspace Linear Delta Expansion}

The purpose of this section is a warming-up with a general presentation of the linear delta expansion (LDE) in the frame of (matter) supersymmetric field theories. We adopt a superfield approach and follow references \cite{WZdelta,OFTdelta,renormOR}. Building up our superspace action in terms of chiral and antichiral supermultiplets, we start off from what we call the interpolated Lagrangian ${\cal L}^{\delta}$:
\begin{equation}
{\mathcal L}^{\delta}=\delta{\mathcal L}(\mu,\bar{\mu})+(1-\delta){\mathcal L}_{0}(\mu,\bar{\mu}) \ , \label{LDE} 
\end{equation}
where $\delta$ is an arbitrary parameter, ${\cal L}_0(\mu,\bar{\mu})$ is the free sector of the Lagrangian, and $\mu$, $\bar{\mu}$ 
are mass parameters. Notice that, whenever $\delta =1$, we recover the original Lagrangian. The $\delta$ parameter 
appears in connection with the interaction terms and is so chosen to be the perturbative expansion parameter; this means that we do not
perturbatively expand in terms of the coupling constant itself. The mass parameters appear in ${\cal L}_0$ and  $\delta{\cal L}_0$. The $(\mu,\bar{\mu})$ dependence of ${\cal L}_0$ is summed up into the propagators, whereas $\delta{\cal L}_0$ is regarded as an insertion and is taken as a quadratic interaction. 
 
Let us now state our methodology. We carry out a usual perturbative expansion in $\delta$ and, at the very end 
of the calculations, we take $\delta =1$. At this stage of our approach, ordinary perturbation theory is applied and a finite 
number of Feynman diagrams is calculated; the results are essentially perturbative. However, quantities evaluated at a finite order in 
$\delta$ explicitly depend on the parameters $\mu$ and $\bar{\mu}$. Therefore, it is necessary to fix them up. To do this, we  adopt
the principle of minimal sensitivity (PMS) \cite{PMS}. In this framework,  the effective potential
${\cal V}^{(k)}_{eff}(\mu,\bar{\mu})$,  perturbatively evaluated to order $\delta ^k$, must be taken at 
a point where it is less sensitive to the parameters $\mu$, $\bar{\mu}$. Invoking the PMS, $\mu={\mu_0}$ and 
$\bar{\mu}={\bar{\mu}_0}$ appear as solutions to the equations
\begin{eqnarray}
\left.\frac{\partial{\cal V}^{(k)}_{eff}(\mu,\bar{\mu})}{\partial\mu}\right|_{\mu=\mu_0,\delta=1}&=&0 \ , \nonumber \\ 
\left.\frac{\partial{\cal V}^{(k)}_{eff}(\mu,\bar{\mu})}{\partial\bar{\mu}}\right|_{\bar{\mu}=\bar{\mu}_0,\delta=1}
&=&0 \ , \label{PMS}
\end{eqnarray}
and will come out as functions of the original coupling and fields. We then insert  ${\mu_0}$, $\bar{\mu}_0$ in the expression for our effective potential ${\cal V}^{(k)}_{eff}$, obtaining a nonperturbative result, once our propagators depend on $\mu$, $\bar{\mu}$.

Our method is based upon the calculation of all the diagrams up to a given order in $\delta$, including the vacuum diagrams. In ordinary quantum field theory, for the calculation of the effective potential, we do not in general worry about vacuum diagrams, since they do not depend on the fields. In our approach, the vacuum super-diagrams do depend on $\mu$ and $\bar{\mu}$ and become important to the LDE, once the arbitrary mass parameters will now depend on fields by virtue of our optimization procedure. Thus, in the LDE, an order-by-order calculation of the vacuum diagrams becomes mandatory. On the other hand, vacuum diagrams in superspace are identically zero, due to the Berezin integrals. In view of this undesirable feature, we take, from the onset, the parameters $\mu$, $\bar{\mu}$ as constant superfields and we keep the vacuum superdiagrams till the end of the optimization procedure. To render  the procedure clearer, let us give below the expression for the superfield generating functional in the presence of the sources $J$ and $\bar J$ (chiral and antichiral, respectively):
\begin{equation}
\tilde{Z}[J,\bar{J}]=exp\left[iS_{int}\left(\frac{1}{i}\frac{\delta}{\delta J},\frac{1}{i}\frac{\delta}{\delta\bar{J}}
\right)\right]exp\left[\frac{i}{2}
(J,\bar{J})G^{(M,\bar{M})}\left( \begin{array}{c}
J \\
\bar{J}
\end{array}
\right)\right] \ ,
\end{equation}
with $m$ being the original mass, $M=m+\mu$ and $\bar{M}=m+\bar{\mu}$. $G^{(M,\bar{M})}$ is the matrix form of the propagator and, in addition to the original interaction terms, one has new bilinear chiral and antichiral interaction terms proportional to $\delta\mu$ and $\delta\bar{\mu}$. We are then lead to the following expression for the superfield effective action:
\begin{equation}
\Gamma[\Phi,\bar{\Phi}]=-\frac{i}{2}\ln[sDet(G^{(M,\bar{M})})]-i\ln\tilde{Z}[J,\bar{J}]-\int\!d^{6}zJ(z)\Phi(z)-
\int\!d^{6}\bar{z}\bar{J}(z)\bar{\Phi}(z) \ , \label{sea}
\end{equation}
where $sDet(G^{(M,\bar{M})})$ is the superdeterminant of $G^{(M,\bar{M})}$. It is in general equal to one, but here we keep it, since $G^{(M,\bar{M})}$ depends on $\mu$ and $\bar{\mu}$. Besides, in view of the $\mu$ and $\bar{\mu}$ dependence, the generating functional of the vacuum diagrams, $\tilde{Z}[0,0]$, is not identically one. We can define a normalized functional generator as $Z_N = \frac{\tilde{Z}[J,\bar J]}{\tilde{Z}[0,0]}$, and set the effective action as written in the expression below:
\begin{equation}
\Gamma[\Phi,\bar{\Phi}]=-\frac{i}{2}\ln[sDet(G)]-i\ln\tilde{Z}[J_{0},\bar{J}_{0}]+\Gamma_{N}[\Phi,\bar{\Phi}] \ , 
\label{Gamma Phi barPhi}
\end{equation}
where the sources $J_0$ and $\bar{J}_0$ are defined by the equations 
\begin{eqnarray}
\frac{\delta W[J,\bar{J}]}{\delta J(z)}|_{J=J_{0}}=\frac{\delta W[J,\bar{J}]}{\delta \bar{J}(z)}|_{\bar{J}=\bar{J}_{0}}
=\frac{\delta\tilde{Z}[J,\bar{J}]}{\delta J(z)}|_{J=J_{0}}=\frac{\delta\tilde{Z}[J,\bar{J}]}{\delta\bar{J}(z)}|_{\bar{J}
=\bar{J}_{0}}=0 \ . \label{gerador}
\end{eqnarray}
In (\ref{Gamma Phi barPhi}), the first two terms (usually equal to zero) stand for the vacuum diagrams and $\Gamma_N[\Phi,\bar{\Phi}]$ is the ordinary contribution to the effective action. 
 
Let us now work out the interpolated Lagrangian and the new super-Feynman rules for the O'Raifeartaigh model. The minimal O'Raifeartaigh model is specified by the Lagrangian:
\begin{equation}
{\mathcal L}=\int d^{4}\theta\bar{\Phi}_{i}\Phi_{i}-\left[\int d^{2}\theta\left(\xi\Phi_{0}+m\Phi_{1}\Phi_{2}
+g\Phi_{0}\Phi_{1}^{2}\right)+h.c.\right] \ , \label{O'Raifeartaigh}
\end{equation} 
where $i=0,1,2$.
  
Following the work of Ref. \cite{OFTdelta}, to account for the nonperturbative contributions of all fields in the model, we apply the LDE with the matrix mass parameters $\mu_{ij}$ and $\bar{\mu}_{ij}$. By adding and subtracting these mass terms to a genral O'Raifeartaigh-type action, we obtain
\begin{equation}
{\mathcal L}(\mu,\bar{\mu})={\mathcal L}_{0}(\mu,\bar{\mu})+{\mathcal L}_{int}(\mu,\bar{\mu}) \ ,
\end{equation}
with
\begin{eqnarray} 
{\mathcal L}_{0}(\mu,\bar{\mu})&=&\int d^{4}\theta\bar{\Phi}_{i}\Phi_{i}-\left[\int d^{2}\theta\left(\xi_{i}\Phi_{i}
+\frac{1}{2}M_{ij}\Phi_{i}\Phi_{j}\right)+h.c.\right] \ , \\
{\mathcal L}_{int}(\mu,\bar{\mu})&=&-\left[\int d^{2}\theta\left(\frac{1}{3!}g_{ijk}\Phi_{i}\Phi_{j}\Phi_{k}
-\frac{1}{2}\mu_{ij}\Phi_{i}\Phi_{j}\right)+h.c.\right] \ ,  
\end{eqnarray}
where $M_{ij}=m_{ij}+ \mu_{ij}$ and $i,j,k=0,1,2$ are symmetrized indices.

We now cast the superfield expansions for the arbitrary mass parameters as follows:
\begin{equation}
\mu_{ij}=\lambda_{ijk}\varphi_k=\lambda_{ijk}(\rho_k+\theta^2\chi_k)=\lambda_{ijk}\rho_k+\lambda_{ijk}
\chi_k\theta^2=\rho_{ij}+b_{ij}\theta^2 \ ,
\end{equation}
so that
\begin{equation}
M_{ij}=m_{ij}+\mu_{ij}=(m_{ij}+\rho_{ij})+b_{ij}\theta^2=a_{ij}+b_{ij}\theta^2 \ .
\end{equation}

The interpolated Lagrangian (\ref{LDE}) takes the form
\begin{equation}
{\mathcal L}^{\delta}= {\mathcal L}_{0}^{\delta} + {\mathcal L}_{int}^{\delta} \ ,
\end{equation}
where the free and the interaction Lagrangians are given by: 
\begin{equation}
{\mathcal L}_{0}^{\delta}=\int d^{4}\theta\bar{\Phi}_{i}\Phi_{i}-\left[\int d^{2}\theta\left(\xi_{i}\Phi_{i}
+\frac{1}{2}a_{ij}\Phi_{i}\Phi_{j}+\frac{1}{2}b_{ij}\theta^{2}\Phi_{i}\Phi_{j}\right)+h.c.\right] \ ,
\end{equation}
\begin{equation}
{\mathcal L}_{int}^{\delta}=-\left[\int d^{2}\theta\left(\frac{\delta}{3!}g_{ijk}\Phi_{i}\Phi_{j}\Phi_{k}
-\frac{\delta}{2}\mu_{ij}\Phi_{i}\Phi_{j}\right)+h.c.\right] \ .
\end{equation}

Notice that the interaction Lagrangian now displays soft SUSY-breaking terms proportional to the $\mu$ components. We treat these terms perturbatively in $\delta$ as for normal interaction terms.

Now, to recover the so-called minimal O'Raifeartaigh model, built up in terms of three superfields, with $\delta=1$ (\ref{O'Raifeartaigh}), we choose:
\begin{equation}
\left\{\begin{array}{lllll}
          \xi_{0}=\xi \ ; \\
          M_{01}=a_{01}=\rho_{01}=a \ ; \\
          M_{11}=b_{11}\theta^{2}=b\theta^{2} \ ; \\
          M_{12}=a_{12}=m_{12}+\rho_{12}=m+\rho=M \ ; \\
          g_{011}=g \ ,
          \end{array}\right. \label{choice}
\end{equation}
and all other $\xi_{i}$ and $M_{ij}$ set to zero. With these choices, we obtain
\begin{eqnarray}
{\mathcal L}_{0}^{\delta}&=&\int d^{4}\theta\bar{\Phi}_{i}\Phi_{i}-\left[\int d^{2}\theta\left(\xi\Phi_{0}
+M\Phi_{1}\Phi_{2}+a\Phi_{0}\Phi_{1}+\frac{1}{2}b\theta^{2}\Phi_{1}^{2}\right)+h.c.\right] \ , \nonumber\\
{\mathcal L}_{int}^{\delta}&=&-\left[\int d^{2}\theta\left(\delta g\Phi_{0}\Phi_{1}^{2}-\delta\rho\Phi_{1}
\Phi_{2}-\delta a\Phi_{0}\Phi_{1}-\frac{\delta}{2}b\theta^{2}\Phi_{1}^{2}\right)+h.c.\right] \ .
\end{eqnarray}

This O'Raifeartaigh model possesses an (Abelian) {\it R} symmetry. The {\it R} charges assignments of the chiral superfields $\Phi_0$, $\Phi_1$, $\Phi_2$ are  respectively $R_0= 2$, $R_1= 0$ and $R_2= 2$. To keep the {\it R} symmetry of the interpolated Lagrangian, the {\it R} charges of the parameters $a$ and $b$ are $R_a=0$ and $R_b=0$; they are to be left the same after the optimization procedure. 

The new set of modified propagators can be read off from the free Lagrangian, which has also explicit dependence on $\theta$ and $\bar{\theta}$ through $\mu$ and $\bar{\mu}$. Using the techniques developed in \cite{Helayel}, the new superfield propagators read as follows:
\begin{eqnarray}
\langle\Phi_{0}\bar{\Phi}_{0}\rangle&=&(k^{2}+|M|^{2})A(k)\delta^{4}_{12}+|a|^{2}|b|^{2}B(k)
\theta_{1}^{2}\bar{\theta}_{1}^{2}\delta_{12}^{4} \ ; \nonumber\\
\langle\Phi_{0}\bar{\Phi}_{1}\rangle&=&\bar{a}bC(k)\frac{1}{16}D_{1}^{2}\bar{D}_{1}^{2}\theta_{1}^{2}
\delta_{12}^{4} \ ; \nonumber\\
\langle\Phi_{0}\bar{\Phi}_{2}\rangle&=&-M\bar{a}A(k)\delta^{4}_{12}+M\bar{a}|b|^{2}B(k)\theta_{1}^{2}
\bar{\theta}_{1}^{2}\delta^{4}_{12} \ ; \nonumber\\
\langle\Phi_{1}\bar{\Phi}_{1}\rangle&=&E(k)\delta_{12}^{4}+|b|^{2}B(k)\frac{1}{16}D_{1}^{2}
\theta_{1}^{2}\bar{\theta}_{1}^{2}\bar{D}_{1}^{2}\delta_{12}^{4} \ ; \nonumber\\
\langle\Phi_{1}\bar{\Phi}_{2}\rangle&=&-M\bar{b}F(k)\bar{\theta}_{1}^{2}\delta^{4}_{12} \ ; \nonumber\\
\langle\Phi_{2}\bar{\Phi}_{2}\rangle&=&(k^{2}+|a|^{2})A(k)\delta^{4}_{12}+|M|^{2}|b|^{2}B(k)
\theta_{1}^{2}\bar{\theta}_{1}^{2}\delta_{12}^{4} \ ; \nonumber\\
\langle\Phi_{0}\Phi_{0}\rangle&=&-|a|^{2}\bar{b}C(k)\frac{1}{4}D_{1}^{2}\theta_{1}^{2}\delta_{12}^{4} \ ; \nonumber\\
\langle\Phi_{0}\Phi_{1}\rangle&=&\bar{a}A(k)\frac{1}{4}D_{1}^{2}\delta_{12}^{4}-\bar{a}|b|^{2}B(k)\frac{1}{4}\theta_{1}^{2}
\bar{\theta}_{1}^{2}D_{1}^{2}\delta_{12}^{4} \ ; \nonumber\\
\langle\Phi_{0}\Phi_{2}\rangle&=&-\bar{M}\bar{a}\bar{b}C(k)\frac{1}{4}D_{1}^{2}\theta_{1}^{2}\delta^{4}_{12} \ ; \nonumber\\
\langle\Phi_{1}\Phi_{1}\rangle&=&\bar{b}F(k)\frac{1}{4}\bar{\theta}_{1}^{2}D_{1}^{2}\delta_{12}^{4} \ ; \nonumber\\
\langle\Phi_{1}\Phi_{2}\rangle&=&\bar{M}A(k)\frac{1}{4}D_{1}^{2}\delta^{4}_{12}-\bar{M}|b|^{2}B(k)\frac{1}{4}D_{1}^{2}
\theta_{1}^{2}\bar{\theta}_{1}^{2}\delta^{4}_{12} \ ; \nonumber\\
\langle\Phi_{2}\Phi_{2}\rangle&=&-|M|^{2}\bar{b}C(k)\frac{1}{4}D_{1}^{2}\theta_{1}^{2}\delta^{4}_{12} \ , \label{propagators}
\end{eqnarray}
where
\begin{eqnarray}
A(k)&=&\frac{1}{k^{2}\left(k^{2}+|M|^{2}+|a|^{2}\right)} \ \ , \\ B(k)&=&\frac{1}{\left(k^{2}+|M|^{2}
+|a|^{2}\right)\left[\left(k^{2}+|M|^{2}+|a|^{2}\right)^{2}-|b|^{2}\right]} \ \ , \nonumber\\
C(k)&=&\frac{1}{k^{2}\left[\left(k^{2}+|M|^{2}+|a|^{2}\right)^{2}-|b|^{2}\right]} \ \ , \nonumber\\ 
E(k)&=&\frac{1}{k^{2}+|M|^{2}+|a|^{2}} \ \ , \nonumber\\ 
F(k)&=&\frac{1}{\left(k^{2}+|M|^{2}+|a|^{2}\right)^{2}-|b|^{2}} \ \ .
\end{eqnarray} 

It is noteworthy to highlight the nontrivial dependence of the propagators on the parameters $a$ and $b$: the optimized parameters are present at the poles of the propagators, as if they were typical mass terms.

The new super-Feynman rules for the vertices are:
\begin{eqnarray}
\Phi_{0}\Phi_{1}^{2} \ \ \hbox{vertex}&:&2\delta g\int d^{4}\theta \ ; \nonumber\\
\Phi_{1}\Phi_{2} \ \ \hbox{vertex}&:&-\delta\rho\int d^{4}\theta \ ; \nonumber\\
\Phi_{0}\Phi_{1} \ \ \hbox{vertex}&:&-\delta a\int d^{4}\theta \ ; \nonumber\\
\Phi_{1}\Phi_{1} \ \ \hbox{vertex}&:&-\frac{\delta b}{2}\int d^{4}\theta\theta^{2} \ . \label{vertices}
\end{eqnarray}

We are now ready to start calculating the perturbative effective potential in powers of $\delta$. To do that, we use the vertex functions defined in the expansion of the effective action and consider the compatible vacuum diagrams. In Ref. \cite{OFTdelta}, after the optimization procedure, it is seen that the order-$\delta^{0}$ contribution accounts for the sum of all one-loop diagrams. It was then possible to derive analytical solutions to the optimization procedure before calculating the superspace and loop momentum integrals. Nevertheless, in second order, we are not able to exhibit analytical solutions; so, we must renormalize the theory before the application of the optimization procedure. To do that, a numeric calculation is required.

\begin{figure}
\begin{eqnarray*}
\begin{picture}(350,5) \thicklines
\put(15,0){\circle{25}}\put(30,-3){+}
\put(55,0){\circle{25}}\put(67,0){\line(50,0){20}}\put(73,4){$\Phi_0$}\put(91,-3){+}
\put(116,0){\circle{25}}\put(123,-3){$\times$}\put(132,0){$\theta^2$}\put(120,14){$\Phi_1$}\put(120,-18)
{$\Phi_1$}\put(145,-3){+}
\put(171,0){\circle{25}}\put(183,0){\line(50,0){20}}\put(189,4){$\Phi_1$}\put(208,-3){+}
\put(233,0){\circle{25}}\put(240.5,-3){$\times$}\put(237,14){$\Phi_0$}\put(237,-18){$\Phi_1$}\put(255,-3){+}
\put(282,0){\circle{25}}\put(289,-3){$\times$}\put(286,14){$\Phi_1$}\put(286,-18){$\Phi_2$}\put(304,-3){+}
\put(321,-3){$h.c.$}
\end{picture}
\end{eqnarray*}
\caption{Effective potential up to the order $\delta^1$}
\label{fig1}
\end{figure}
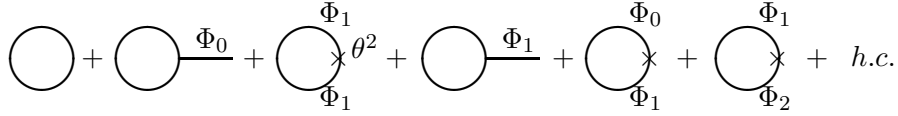

In Fig. \ref{fig1}, we depict the diagrammatic sum of the effective potential up to the order $\delta^{1}$ 
(${\mathcal V}_{eff}^{(1)}$).

Notice that, due to the $\theta$-dependent propagators,  tadpole diagrams do not vanish any longer, as it usually happens in superspace. The first diagram is of order $\delta^{0}$ and we can see that it corresponds to the first term of the effective action, as given in (\ref{Gamma Phi barPhi}). The third graph is a vacuum-type superdiagram with a quadratic insertion that stems from the $\theta^2$ component of the $\mu$ expansion.

With our super-Feynman rules and the results of \cite{OFTdelta}, the effective potential up to the order $\delta^1$ 
reads as given below:
\begin{eqnarray}
{\mathcal V}_{eff}^{(1)}&=&{\mathcal G}^{(1)\delta^{0}}+\sum_{i=1}^{5}{\mathcal G}_{i}^{(1)\delta^{1}} \nonumber\\
&=&\frac{1}{2}\int d^{4}\theta_{12}\delta^{4}_{12}Tr\ln[P^{T}K]\delta^{4}_{12}+\delta\int\frac{d^{4}k}{(2\pi)^{4}}F(k)
\left\{-2g\bar{b}\int d^{2}\theta\Phi_{0}+\frac{1}{2}\left|b\right|^{2}+h.c.\right\} \nonumber\\
&&+\delta\int\frac{d^{4}k}{(2\pi)^{4}}B(k)\left\{4g\bar{a}\left|b\right|^{2}\int d^{2}\theta\theta^{2}\Phi_{1}-\left|a\right|^{2}
\left|b\right|^{2}-\rho\bar{M}\left|b\right|^{2}+h.c.\right\} \ . \label{pot efet OR ordem 1}
\end{eqnarray} 

In Ref. \cite{renormOR}, the renormalization of the effective potential up to the order $\delta^2$ is discussed. At the order $\delta^1$, the counterterm below is needed:
\begin{equation}
\frac{2\delta gb}{\kappa\epsilon}\int d^{4}\theta\bar{\theta}^{2}\Phi_{0R}+h.c.=\frac{2\delta gb}
{\kappa\epsilon}\int d^{2}\theta\Phi_{0R}+h.c. \ , \label{CT1}
\end{equation}
where $\Phi_{0R}$ denotes the renormalized superfield. The chiral potential is therefore renormalized. The counterterm that renormalizes it depends on $b$, which is a solution to the optimization procedure. In \cite{renormOR}, it was shown that, after the optimization, only the K\"{a}hler potential is actually renormalized.  

The renormalized effective potential up to the order $\delta^1$ is  
\begin{eqnarray}
{\mathcal V}_{eff}^{(1)}&=&\frac{1}{(4\pi)^{2}}\left\{\frac{1}{4}\!\left(M^{2}\!+\!a^{2}\right)^{2}\ln\!\left[1-\frac{b^{2}}
{\left(M^{2}+a^{2}\right)^{2}}\right]\right. \nonumber\\
&&\left.+\frac{b}{2}\!\left(M^{2}\!+\!a^{2}\right)\!\ln\!\left[\frac{M^{2}+a^{2}+b}{M^{2}+a^{2}-b}\right]\!
+\!\frac{b^{2}}{4}\ln\!\left[\frac{\left(M^{2}\!+\!a^{2}\right)^{2}-b^{2}}{\mu^4}\right]\!-\!\frac{3b^2}{4}\right\} \nonumber\\
&&+\frac{\delta}{(4\pi)^{2}}\!\left\{b(b-4g\langle F_{0}\rangle)+2\left[a(a-4g\langle\varphi_{1}\rangle)+\rho M\right]\!\left(M^{2}
+a^{2}\right)\overline{\ln}\left[M^{2}+a^{2}\right]\right. \nonumber\\
&&\!\left.+\left[a(4g\langle\varphi_{1}\rangle\!-\!a)\!+\!\frac{1}{2}(4g\langle F_{0}\rangle\!-\!b)\!-\!\rho M\right]\!\!\left(M^{2}\!
+\!a^{2}\!+\!b\right)\overline{\ln}\left[M^{2}\!+\!a^{2}\!+\!b\right]\right. \nonumber\\
&&\!\left.+\left[a(4g\langle\varphi_{1}\rangle\!-\!a)\!-\!\frac{1}{2}(4g\langle F_{0}\rangle\!-\!b)\!-\!\rho M\right]\!\!\left(M^{2}\!
+\!a^{2}\!-\!b\right)\overline{\ln}\left[M^{2}\!+\!a^{2}\!-\!b\right]\!\right\}. \label{1loopdelta1}
\end{eqnarray}

For the time being, what we have is a perturbative result for the effective potential. To actually obtain a nonperturbative result, we apply the optimization procedure. We had break the parameters $M_{ij}$ into $\theta$-independent ($a_{ij}$) and $\theta$-dependent ($b_{ij}$) parts and, with the help of  (\ref{choice}), the optimized parameters turn out to be $a_{01}=a$, $b_{11}=b$, and $\rho_{12}=\rho$. Upon application of the PMS, to find the optimized parameters $a$, $b$ and $\rho$, we have to solve the three coupled equations
\begin{equation}
\left.\frac{\partial{\mathcal V}_{eff}^{(1)}}{\partial a}\right|_{a=a_0}=\left.\frac{\partial{\mathcal V}_{eff}^{(1)}}
{\partial b}\right|_{b=b_0}=\left.\frac{\partial{\mathcal V}_{eff}^{(1)}}{\partial\rho}\right|_{\rho=\rho_0}=0 \ , 
\label{PMS OR 2 Veff}
\end{equation}
at $\delta=1$, and plug the optimized values $a_0$, $b_0$ and $\rho_0$ into (\ref{1loopdelta1}). The following analytical solutions are encountered:
\begin{eqnarray}
a_{0}&=&4g\langle\varphi_{1}\rangle=\bar{a}_{0} \nonumber\\
b_{0}&=&4g\langle F_{0}\rangle=\bar{b}_{0} \ , \nonumber \\
\rho_{0}&=&0=\bar{\rho}_{0} \ . \label{analiticalsol}
\end{eqnarray}

The optimized parameters appear now as functions of the original coupling and fields, as expected. By plugging these results into (\ref{1loopdelta1}), all the $\delta^1$ terms vanish and the optimized potential can be cast under the form below:
\begin{eqnarray}
{\mathcal V}_{eff}^{(1)}={\mathcal G}^{(1)\delta^0}&=&\frac{1}{(8\pi)^{2}}\left\{(m^{2}+16g^{2}\langle\varphi_{1}\rangle^{2})^{2}
\ln\left[1-\frac{16g^{2}\langle F_{0}\rangle^{2}}{(m^{2}+16g^{2}\langle\varphi_{1}\rangle^{2})^{2}}\right]\right. \nonumber\\ 
&&\left.+8g\langle F_{0}\rangle(m^{2}+16g^{2}\langle\varphi_{1}\rangle^{2})\ln\left[\frac{m^{2}+16g^{2}\langle\varphi_{1}\rangle^{2}
+4g\langle F_{0}\rangle}{m^{2}+16g^{2}\langle\varphi_{1}\rangle^{2}-4g\langle F_{0}\rangle}\right]\right. \nonumber\\
&&\left.+16g^{2}\langle F_{0}\rangle^{2}\ln\left[\frac{(m^{2}+16g^{2}\langle\varphi_{1}\rangle^{2})^{2}-16g^{2}\langle F_{0}
\rangle^{2}}{\mu^4}\right]- 48g^2\langle F_0\rangle ^2\right\} \ . \label{pot Coleman Weinberg OR final}  
\end{eqnarray}

The above result represents the Coleman-Weinberg-type potential~\cite{CW,Jac} for the O'Raifeartaigh model~\cite{OFTdelta,Helayel} and it accounts for the sum of all one-loop diagrams. It is a nonperturbative result in that it takes into account all orders (actually infinite orders) in the original coupling constant.  


\section{Effective potential in the LDE at $\mathcal{O}(\delta^2)$}

We now present the order-$\delta^{2}$ results. At this order, we have one- and two-loop diagrams. Since the 
purpose of the present work is to be as clear as possible, the  main part of this section is rather technical.  All 
the results of this section are summarized in section IV.

\subsection{One Loop}
To analyze the one-loop diagrams we adopt the following strategy: there are 42 diagrams (plus some hermitian 
conjugates), and we separate them in distinct sets; each diagram belonging to a certain set has the same 
propagator structure in the loop and differs from the other diagrams of the set only by the external classical 
superfields or by insertions of $a$, $b$ or $\rho$. Below, we show the expressions for these diagrams, set by set. 
We are going to use the notation ${\mathcal J}_{i}(\theta,\bar\theta)$ for superspace integrals, which are listed 
in the Appendix. 

The first set is:	
\begin{figure}[hb]
\begin{eqnarray*}
\begin{picture}(335,5) \thicklines 
\put(15,0){\line(50,0){30}}\put(15,4){$\Phi_0$}\put(35,12){$\Phi_1$}\put(35,-17){$\Phi_1$}\put(61,0){\circle{30}}
\put(75,12){$\Phi_1$}\put(75,-17){$\Phi_1$}\put(77,0){\line(50,0){30}}\put(94,4){$\Phi_0$}\put(114,-3){;}
\put(125,0){\line(50,0){30}}\put(125,4){$\Phi_0$}\put(145,12){$\Phi_1$}\put(145,-17){$\Phi_1$}\put(171,0)
{\circle{30}}\put(185,12){$\Phi_1$}\put(185,-17){$\Phi_1$}\put(182,-3){$\times$}\put(192,-3){$\theta^2$}
\put(208,-3){;}
\put(217,-3){$\theta^2$}\put(227,-3){$\times$}\put(222,12){$\Phi_1$}\put(222,-17){$\Phi_1$}\put(248,0)
{\circle{30}}\put(262,12){$\Phi_1$}\put(262,-17){$\Phi_1$}\put(259,-3){$\times$}\put(269,-3){$\theta^2$}
\put(284,-3){;}
\put(293,-3){+ $h.c.$}
\end{picture}
\end{eqnarray*}
\caption{Diagrams ${\mathcal G}_{1}^{(1)\delta^2}$, ${\mathcal G}_{2}^{(1)\delta^2}$ and ${\mathcal G}_{3}^{(1)
\delta^2}$}
\label{fig2}
\end{figure}

\begin{eqnarray}
{\mathcal G}_{1}^{(1)\delta^2}&=&4\delta^{2}g^{2}\int\frac{d^{4}kd^{4}\theta_{12}}{(2\pi)^{4}}\Phi_{0}(1)\Phi_{0}(2)
\left[-\frac{1}{4}\bar{D}_{1}^{2}(k)\langle\Phi_{1}\Phi_{1}\rangle\right]\left[-\frac{1}{4}\bar{D}_{2}^{2}(-k)\langle\Phi_{1}
\Phi_{1}\rangle\right]+h.c. \nonumber\\
&=&\frac{4\delta^{2}g^{2}}{(16)^{2}}\bar{b}^{2}\int\frac{d^{4}k}{(2\pi)^{4}}F(k)F(k){\mathcal J}_{1}(\theta,\bar\theta)
+h.c. \nonumber\\
&=&\frac{2\delta^{2}g^{2}\langle F_{0}\rangle^{2}}{\kappa b}\left(\eta^{2}\overline{\ln}\eta^{+}-\eta^{2}\overline{\ln}
\eta^{-}-2b\right) \ . \label{G1}
\end{eqnarray}
\begin{eqnarray}
{\mathcal G}_{2}^{(1)\delta^2}&=&-2\delta^{2}gb\int\frac{d^{4}kd^{4}\theta_{12}}{(2\pi)^{4}}\Phi_{0}(1)\theta_{2}^{2}
\left[-\frac{1}{4}\bar{D}_{1}^{2}(k)\langle\Phi_{1}\Phi_{1}\rangle\right]\left[-\frac{1}{4}\bar{D}_{2}^{2}(-k)\langle\Phi_{1}\Phi_{1}\rangle\right]+h.c. \nonumber\\
&=&-\frac{2\delta^{2}g}{(16)^{2}}\bar{b}|b|^{2}\int\frac{d^{4}k}{(2\pi)^{4}}F(k)F(k){\mathcal J}_{2}(\theta,\bar\theta)
+h.c. \nonumber\\
&=&-\frac{\delta^{2}gb\langle F_{0}\rangle}{\kappa b}\left(\eta^{2}\overline{\ln}\eta^{+}-\eta^{2}\overline{\ln}\eta^{-}-2b\right) \ . \label{G2}
\end{eqnarray}
\begin{eqnarray}
{\mathcal G}_{3}^{(1)\delta^2}&=&\frac{1}{4}\delta^{2}b^{2}\int\frac{d^{4}kd^{4}\theta_{12}}{(2\pi)^{4}}\theta_{1}^{2}
\theta_{2}^{2}\left[-\frac{1}{4}\bar{D}_{1}^{2}(k)\langle\Phi_{1}\Phi_{1}\rangle\right]\left[-\frac{1}{4}\bar{D}_{2}^{2}(-k)\langle\Phi_{1}\Phi_{1}\rangle\right]+h.c. \nonumber\\
&=&\frac{\delta^{2}}{4(16)^{2}}|b|^{4}\int\frac{d^{4}k}{(2\pi)^{4}}F(k)F(k){\mathcal J}_{3}(\theta,\bar\theta)+h.c. \nonumber\\
&=&\frac{\delta^{2}b^{2}}{8\kappa b}\left(\eta^{2}\overline{\ln}\eta^{+}-\eta^{2}\overline{\ln}\eta^{-}-2b\right) \ . \label{G3}
\end{eqnarray}

The second set is:

\begin{figure}[htb!]
\begin{eqnarray*}
\begin{picture}(425,5) \thicklines    
\put(15,0){\line(50,0){30}}\put(15,4){$\Phi_0$}\put(35,12){$\Phi_1$}\put(35,-17){$\Phi_1$}\put(61,0){\circle{30}}
\put(75,12){$\Phi_0$}\put(75,-17){$\Phi_1$}\put(77,0){\line(50,0){30}}\put(94,4){$\Phi_1$}\put(114,-3){;}
\put(125,0){\line(50,0){30}}\put(125,4){$\Phi_0$}\put(145,12){$\Phi_1$}\put(145,-17){$\Phi_1$}\put(171,0)
{\circle{30}}\put(185,12){$\Phi_0$}\put(185,-17){$\Phi_1$}\put(182,-3){$\times$}\put(206,-3){;}
\put(218,0){\line(50,0){30}}\put(218,4){$\Phi_1$}\put(238,12){$\Phi_0$}\put(238,-17){$\Phi_1$}\put(264,0)
{\circle{30}}\put(278,12){$\Phi_1$}\put(278,-17){$\Phi_1$}\put(275,-3){$\times$}\put(285,-3){$\theta^2$}
\put(301,-3){;}
\put(311,-3){$\theta^2$}\put(321,-3){$\times$}\put(316,12){$\Phi_1$}\put(316,-17){$\Phi_1$}\put(342,0)
{\circle{30}}\put(356,12){$\Phi_0$}\put(356,-17){$\Phi_1$}\put(353,-3){$\times$}\put(376,-3){;}
\put(385,-3){+ $h.c.$}
\end{picture}
\end{eqnarray*}
\caption{Diagrams ${\mathcal G}_{4}^{(1)\delta^2}$, ${\mathcal G}_{5}^{(1)\delta^2}$, 
${\mathcal G}_{6}^{(1)\delta^2}$ and ${\mathcal G}_{7}^{(1)\delta^2}$. }
\label{fig3}
\end{figure}
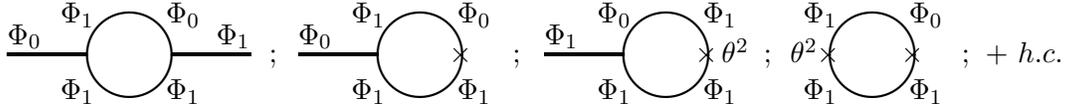

\begin{eqnarray}
{\mathcal G}_{4}^{(1)\delta^2}&=&16\delta^{2}g^{2}\int\frac{d^{4}kd^{4}\theta_{12}}{(2\pi)^{4}}\Phi_{0}(1)\Phi_{1}(2)
\left[-\frac{1}{4}\bar{D}_{1}^{2}(k)\langle\Phi_{1}\Phi_{1}\rangle\right]\left[-\frac{1}{4}\bar{D}_{2}^{2}(-k)\langle\Phi_{0}\Phi_{1}\rangle\right]+h.c. \nonumber\\
&=&\frac{16\delta^{2}g^{2}}{(16)^{2}}\bar{a}\bar{b}\int\frac{d^{4}k}{(2\pi)^{4}}F(k)\left\{A(k){\mathcal J}_{4}(\theta,\bar\theta)
-|b|^{2}B(k){\mathcal J}_{5}(\theta,\bar\theta)\right\}+h.c. \nonumber\\
&=&-\frac{8\delta^{2}g^{2}a\langle F_{0}\rangle\langle\varphi_{1}\rangle}{\kappa}\left(\overline{\ln}\eta^{+}-\overline{\ln}\eta^{-}\right) \ . \label{G4}
\end{eqnarray}

\begin{eqnarray}
{\mathcal G}_{5}^{(1)\delta^2}&=&-4\delta^{2}ga\int\frac{d^{4}kd^{4}\theta_{12}}{(2\pi)^{4}}\Phi_{0}(1)\left[-\frac{1}{4}
\bar{D}_{1}^{2}(k)\langle\Phi_{1}\Phi_{1}\rangle\right]\left[-\frac{1}{4}\bar{D}_{2}^{2}(-k)\langle\Phi_{0}\Phi_{1}\rangle\right]+h.c. \nonumber\\
&=&-\frac{4\delta^{2}g}{(16)^{2}}|a|^{2}\bar{b}\int\frac{d^{4}k}{(2\pi)^{4}}F(k)\left\{A(k){\mathcal J}_{6}(\theta,\bar\theta)
-|b|^{2}B(k){\mathcal J}_{2}(\theta,\bar\theta)\right\}+h.c. \nonumber\\
&=&\frac{2\delta^{2}ga^{2}\langle F_{0}\rangle}{\kappa}\left(\overline{\ln}\eta^{+}-\overline{\ln}\eta^{-}\right) \ . \label{G5}
\end{eqnarray}
\begin{eqnarray}
{\mathcal G}_{6}^{(1)\delta^2}&=&-4\delta^{2}gb\int\frac{d^{4}kd^{4}\theta_{12}}{(2\pi)^{4}}\Phi_{1}(1)\theta_{2}^{2}
\left[-\frac{1}{4}\bar{D}_{1}^{2}(k)\langle\Phi_{1}\Phi_{1}\rangle\right]\left[-\frac{1}{4}\bar{D}_{2}^{2}(-k)\langle\Phi_{1}\Phi_{0}\rangle\right]+h.c. \nonumber\\
&=&-\frac{4\delta^{2}g}{(16)^{2}}\bar{a}|b|^{2}\int\frac{d^{4}k}{(2\pi)^{4}}F(k)\left\{A(k){\mathcal J}_{7}(\theta,\bar\theta)-|b|^{2}
B(k){\mathcal J}_{8}(\theta,\bar\theta)\right\}+h.c. \nonumber\\
&=&\frac{2\delta^{2}gab\langle\varphi_{1}\rangle}{\kappa}\left(\overline{\ln}\eta^{+}-\overline{\ln}\eta^{-}\right) \ . \label{G6}
\end{eqnarray}
\begin{eqnarray}
{\mathcal G}_{7}^{(1)\delta^2}&=&\delta^{2}ab\int\frac{d^{4}kd^{4}\theta_{12}}{(2\pi)^{4}}\theta_{1}^{2}\left[-\frac{1}{4}
\bar{D}_{1}^{2}(k)\langle\Phi_{1}\Phi_{1}\rangle\right]\left[-\frac{1}{4}\bar{D}_{2}^{2}(-k)\langle\Phi_{0}\Phi_{1}\rangle\right]+h.c. \nonumber\\
&=&\frac{\delta^{2}}{(16)^{2}}|a|^{2}|b|^{2}\int\frac{d^{4}k}{(2\pi)^{4}}F(k)\left\{A(k){\mathcal J}_{9}(\theta,\bar\theta)
-|b|^{2}B(k){\mathcal J}_{3}(\theta,\bar\theta)\right\}+h.c. \nonumber\\
&=&-\frac{a^{2}b}{2\kappa}\left(\overline{\ln}\eta^{+}-\overline{\ln}\eta^{-}\right) \ . \label{G7}
\end{eqnarray}

The third set is:

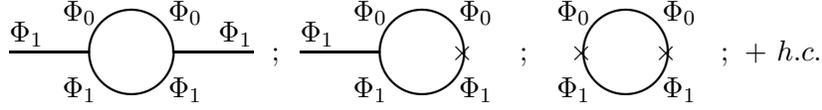
\begin{figure}[htb!]
\begin{eqnarray*}
\begin{picture}(335,5) \thicklines 
\put(15,0){\line(50,0){30}}\put(15,4){$\Phi_1$}\put(35,12){$\Phi_0$}\put(35,-17){$\Phi_1$}\put(61,0){\circle{30}}\put(75,12){$\Phi_0$}\put(75,-17){$\Phi_1$}\put(77,0){\line(50,0){30}}\put(94,4){$\Phi_1$}\put(114,-3){;}
\put(125,0){\line(50,0){30}}\put(125,4){$\Phi_1$}\put(145,12){$\Phi_0$}\put(145,-17){$\Phi_1$}\put(171,0){\circle{30}}\put(185,12){$\Phi_0$}\put(185,-17){$\Phi_1$}\put(182,-3){$\times$}\put(208,-3){;}
\put(227,-3){$\times$}\put(222,12){$\Phi_0$}\put(222,-17){$\Phi_1$}\put(248,0){\circle{30}}\put(262,12){$\Phi_0$}\put(262,-17){$\Phi_1$}\put(259,-3){$\times$}\put(284,-3){;}
\put(293,-3){+ $h.c.$}
\end{picture}
\end{eqnarray*}
\caption{Diagrams ${\mathcal G}_{8}^{(1)\delta^2}$, ${\mathcal G}_{9}^{(1)\delta^2}$ and ${\mathcal G}_{10}^{(1)\delta^2}$. }
\label{fig4}
\end{figure}

\begin{eqnarray}
{\mathcal G}_{8}^{(1)\delta^2}&=&8\delta^{2}g^{2}\int\frac{d^{4}kd^{4}\theta_{12}}{(2\pi)^{4}}\Phi_{1}(1)\Phi_{1}(2)\left[-\frac{1}{4}\bar{D}_{1}^{2}(k)\langle\Phi_{1}\Phi_{1}\rangle\right]\left[-\frac{1}{4}\bar{D}_{2}^{2}(-k)\langle\Phi_{0}\Phi_{0}\rangle\right]+h.c. \nonumber\\
&=&-\frac{8\delta^{2}g^{2}}{(16)^{2}}|a|^{2}\bar{b}^{2}\int\frac{d^{4}k}{(2\pi)^{4}}F(k)C(k){\mathcal J}_{10}(\theta,\bar\theta)+h.c. \nonumber\\
&=&\frac{4\delta^{2}g^{2}a^{2}\langle\varphi_{1}\rangle^{2}}{\kappa b}\left(\eta^{2}\overline{\ln}\eta^{+}-\eta^{2}\overline{\ln}\eta^{-}-2b\right) \ . \label{G8}
\end{eqnarray}
\begin{eqnarray}
{\mathcal G}_{9}^{(1)\delta^2}&=&-4\delta^{2}ga\int\frac{d^{4}kd^{4}\theta_{12}}{(2\pi)^{4}}\Phi_{1}(1)\left[-\frac{1}{4}\bar{D}_{1}^{2}(k)\langle\Phi_{1}\Phi_{1}\rangle\right]\left[-\frac{1}{4}\bar{D}_{2}^{2}(-k)\langle\Phi_{0}\Phi_{0}\rangle\right]+h.c. \nonumber\\
&=&\frac{4\delta^{2}g}{(16)^{2}}a|a|^{2}\bar{b}^{2}\int\frac{d^{4}k}{(2\pi)^{4}}F(k)C(k){\mathcal J}_{7}(\theta,\bar\theta)+h.c. \nonumber\\
&=&-\frac{2\delta^{2}ga^{3}\langle\varphi_{1}\rangle}{\kappa b}\left(\eta^{2}\overline{\ln}\eta^{+}-\eta^{2}\overline{\ln}\eta^{-}-2b\right) \ . \label{G9}
\end{eqnarray}
\begin{eqnarray}
{\mathcal G}_{10}^{(1)\delta^2}&=&\frac{1}{2}\delta^{2}a^{2}\int\frac{d^{4}kd^{4}\theta_{12}}{(2\pi)^{4}}\left[-\frac{1}{4}\bar{D}_{1}^{2}(k)\langle\Phi_{1}\Phi_{1}\rangle\right]\left[-\frac{1}{4}\bar{D}_{2}^{2}(-k)\langle\Phi_{0}\Phi_{0}\rangle\right]+h.c. \nonumber\\
&=&-\frac{\delta^{2}}{2(16)^{2}}a^{2}|a|^{2}\bar{b}^{2}\int\frac{d^{4}k}{(2\pi)^{4}}F(k)C(k){\mathcal J}_{9}(\theta,\bar\theta)+h.c. \nonumber\\
&=&\frac{\delta^{2}a^{4}}{4\kappa b}\left(\eta^{2}\overline{\ln}\eta^{+}-\eta^{2}\overline{\ln}\eta^{-}-2b\right) \ . \label{G10}
\end{eqnarray}
\newpage
The fourth set is:

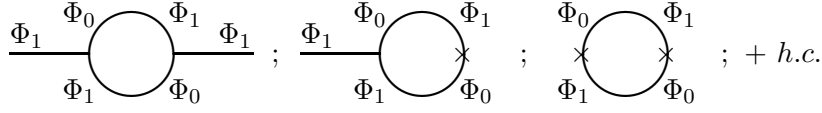
\begin{figure}[htb]
\begin{eqnarray*}
\begin{picture}(335,5) \thicklines 
\put(15,0){\line(50,0){30}}\put(15,4){$\Phi_1$}\put(35,12){$\Phi_0$}\put(35,-17){$\Phi_1$}\put(61,0){\circle{30}}\put(75,12){$\Phi_1$}\put(75,-17){$\Phi_0$}\put(77,0){\line(50,0){30}}\put(94,4){$\Phi_1$}\put(114,-3){;}
\put(125,0){\line(50,0){30}}\put(125,4){$\Phi_1$}\put(145,12){$\Phi_0$}\put(145,-17){$\Phi_1$}\put(171,0){\circle{30}}\put(185,12){$\Phi_1$}\put(185,-17){$\Phi_0$}\put(182,-3){$\times$}\put(208,-3){;}
\put(227,-3){$\times$}\put(222,12){$\Phi_0$}\put(222,-17){$\Phi_1$}\put(248,0){\circle{30}}\put(262,12){$\Phi_1$}\put(262,-17){$\Phi_0$}\put(259,-3){$\times$}\put(284,-3){;}
\put(293,-3){+ $h.c.$}
\end{picture}
\end{eqnarray*}
\caption{Diagrams ${\mathcal G}_{11}^{(1)\delta^2}$, ${\mathcal G}_{12}^{(1)\delta^2}$ and ${\mathcal G}_{13}^{(1)\delta^2}$.}
\label{fig5}
\end{figure}

\begin{eqnarray}
{\mathcal G}_{11}^{(1)\delta^2}&=&8\delta^{2}g^{2}\int\frac{d^{4}kd^{4}\theta_{12}}{(2\pi)^{4}}\Phi_{1}(1)\Phi_{1}(2)\left[-\frac{1}{4}\bar{D}_{1}^{2}(k)\langle\Phi_{1}\Phi_{0}\rangle\right]\left[-\frac{1}{4}\bar{D}_{2}^{2}(-k)\langle\Phi_{1}\Phi_{0}\rangle\right]+h.c. \nonumber\\
&=&\frac{8\delta^{2}g^{2}}{(16)^{2}}\bar{a}^{2}\int\frac{d^{4}k}{(2\pi)^{4}}\left\{A(k)A(k){\mathcal J}_{11}(\theta,\bar\theta)-2|b|^{2}A(k)B(k){\mathcal J}_{10}(\theta,\bar\theta)\right. \nonumber\\
&&\left.+|b|^{4}B(k)B(k){\mathcal J}_{12}(\theta,\bar\theta)\right\}+h.c. \nonumber\\
&=&\frac{4\delta^{2}g^{2}a^{2}\langle\varphi_{1}\rangle^{2}}{\kappa b}\left[4b\overline{\ln}\eta^{2}-(\eta^{2}+2b)\overline{\ln}\eta^{+}+(\eta^{2}-2b)\overline{\ln}\eta^{-}+2b\right] \ . \label{G11}
\end{eqnarray}
\begin{eqnarray}
{\mathcal G}_{12}^{(1)\delta^2}&=&-4\delta^{2}ga\int\frac{d^{4}kd^{4}\theta_{12}}{(2\pi)^{4}}\Phi_{1}(1)\left[-\frac{1}{4}\bar{D}_{1}^{2}(k)\langle\Phi_{1}\Phi_{0}\rangle\right]\left[-\frac{1}{4}\bar{D}_{2}^{2}(-k)\langle\Phi_{1}\Phi_{0}\rangle\right]+h.c. \nonumber\\
&=&-\frac{4\delta^{2}g}{(16)^{2}}|a|^{2}\bar{a}\int\frac{d^{4}k}{(2\pi)^{4}}\left\{A(k)A(k){\mathcal J}_{13}(\theta,\bar\theta)-2|b|^{2}A(k)B(k){\mathcal J}_{7}(\theta,\bar\theta)\right. \nonumber\\
&&\left.+|b|^{4}B(k)B(k){\mathcal J}_{8}(\theta,\bar\theta)\right\}+h.c. \nonumber\\
&=&-\frac{2\delta^{2}ga^{3}\langle\varphi_{1}\rangle}{\kappa b}\left[4b\overline{\ln}\eta^{2}-(\eta^{2}+2b)\overline{\ln}\eta^{+}+(\eta^{2}-2b)\overline{\ln}\eta^{-}+2b\right] \ . \label{G12}
\end{eqnarray}
\begin{eqnarray}
{\mathcal G}_{13}^{(1)\delta^2}&=&\frac{1}{2}\delta^{2}a^{2}\int\frac{d^{4}kd^{4}\theta_{12}}{(2\pi)^{4}}\left[-\frac{1}{4}\bar{D}_{1}^{2}(k)\langle\Phi_{1}\Phi_{0}\rangle\right]\left[-\frac{1}{4}\bar{D}_{2}^{2}(-k)\langle\Phi_{1}\Phi_{0}\rangle\right]+h.c. \nonumber\\
&=&\frac{\delta^{2}}{2(16)^{2}}|a|^{4}\int\frac{d^{4}k}{(2\pi)^{4}}\left\{A(k)A(k){\mathcal J}_{14}(\theta,\bar\theta)-2|b|^{2}A(k)B(k){\mathcal J}_{9}(\theta,\bar\theta)\right. \nonumber\\
&&\left.+|b|^{4}B(k)B(k){\mathcal J}_{3}(\theta,\bar\theta)\right\}+h.c. \nonumber\\
&=&\frac{\delta^{2}a^{4}}{4\kappa b}\left[4b\overline{\ln}\eta^{2}-(\eta^{2}+2b)\overline{\ln}\eta^{+}+(\eta^{2}-2b)\overline{\ln}\eta^{-}+2b\right] \ . \label{G13}
\end{eqnarray}

The fifth set is:
\begin{figure}[htb!]
\begin{eqnarray*}
\begin{picture}(330,5) \thicklines 
\put(15,0){\line(50,0){30}}\put(15,4){$\Phi_0$}\put(35,12){$\Phi_1$}\put(35,-17){$\Phi_1$}\put(61,0){\circle{30}}\put(75,12){$\bar{\Phi}_1$}\put(75,-17){$\bar{\Phi}_1$}\put(77,0){\line(50,0){30}}\put(94,4){$\bar{\Phi}_0$}\put(114,-3){;}
\put(125,0){\line(50,0){30}}\put(125,4){$\Phi_0$}\put(145,12){$\Phi_1$}\put(145,-17){$\Phi_1$}\put(171,0){\circle{30}}\put(185,12){$\bar{\Phi}_1$}\put(185,-17){$\bar{\Phi}_1$}\put(182,-3){$\otimes$}\put(192,-3){$\bar{\theta}^2$}\put(208,-3){+ $h.c.$}\put(245,-3){;}
\put(255,-3){$\theta^2$}\put(265,-3){$\times$}\put(260,12){$\Phi_1$}\put(260,-17){$\Phi_1$}\put(286,0){\circle{30}}\put(300,12){$\bar{\Phi}_1$}\put(300,-17){$\bar{\Phi}_1$}\put(297,-3){$\otimes$}\put(307,-3){$\bar{\theta}^2$}
\end{picture}
\end{eqnarray*}
\caption{Diagrams ${\mathcal G}_{14}^{(1)\delta^2}$, ${\mathcal G}_{15}^{(1)\delta^2}$ and ${\mathcal G}_{16}^{(1)\delta^2}$.}
\label{fig6}
\end{figure}
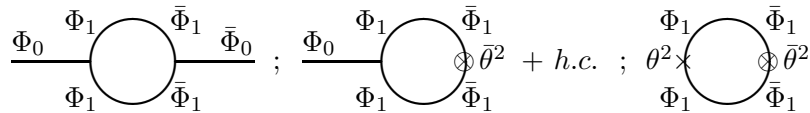

\begin{eqnarray}
{\mathcal G}_{14}^{(1)\delta^2}&=&8\delta^{2}g^{2}\int\frac{d^{4}kd^{4}\theta_{12}}{(2\pi)^{4}}\Phi_{0}(1)\bar{\Phi}_{0}(2)\left[-\frac{1}{4}\bar{D}_{1}^{2}(k)\langle\Phi_{1}\bar{\Phi}_{1}\rangle\right]\left[-\frac{1}{4}D_{2}^{2}(-k)\langle\bar{\Phi}_{1}\Phi_{1}\rangle\right] \nonumber\\
&=&\frac{8\delta^{2}g^{2}}{16}\int\frac{d^{4}k}{(2\pi)^{4}}\left\{E(k)E(k){\mathcal J}_{15}(\theta,\bar\theta)+2\frac{|b|^{2}}{16}E(k)B(k){\mathcal J}_{16}(\theta,\bar\theta)\right. \nonumber\\
&&\left.+\frac{|b|^{4}}{(16)^{2}}B(k)B(k){\mathcal J}_{17}(\theta,\bar\theta)\right\} \nonumber\\
&=&\frac{8\delta^{2}g^{2}\langle F_{0}\rangle^{2}}{\kappa}\left(\frac{1}{\epsilon}-\overline{\ln}\eta^{2}\right) \nonumber\\
&&+\frac{2\delta^{2}g^{2}\langle F_{0}\rangle^{2}}{\kappa b}\left[4b\overline{\ln}\eta^{2}-(\eta^{2}+2b)\overline{\ln}\eta^{+}+(\eta^{2}-2b)\overline{\ln}\eta^{-}+2b\right] \ . \label{G14}
\end{eqnarray}
\begin{eqnarray}
{\mathcal G}_{15}^{(1)\delta^2}&=&-2\delta^{2}g\bar{b}\int\frac{d^{4}kd^{4}\theta_{12}}{(2\pi)^{4}}\Phi_{0}(1)\bar{\theta}_{2}^{2}\left[-\frac{1}{4}
\bar{D}_{1}^{2}(k)\langle\Phi_{1}\bar{\Phi}_{1}\rangle\right]\left[-\frac{1}{4}D_{2}^{2}(-k)\langle\bar{\Phi}_{1}\Phi_{1}\rangle\right]+h.c. \nonumber\\
&=&-\frac{2\delta^{2}g}{16}\bar{b}\int\frac{d^{4}k}{(2\pi)^{4}}\left\{E(k)E(k){\mathcal J}_{18}(\theta,\bar\theta)+2\frac{|b|^{2}}{16}E(k)B(k){\mathcal J}_{2}
(\theta,\bar\theta)\right. \nonumber\\
&&\left.+\frac{|b|^{4}}{(16)^{2}}B(k)B(k){\mathcal J}_{19}(\theta,\bar\theta)\right\}+h.c. \nonumber\\
&=&-\frac{4\delta^{2}gb\langle F_{0}\rangle}{\kappa}\left(\frac{1}{\epsilon}-\overline{\ln}\eta^{2}\right) \nonumber\\
&&-\frac{\delta^{2}gb\langle F_{0}\rangle}{\kappa b}\left[4b\overline{\ln}\eta^{2}-(\eta^{2}+2b)\overline{\ln}\eta^{+}+(\eta^{2}-2b)\overline{\ln}\eta^{-}+2b\right] \ . \label{G15}
\end{eqnarray}
\begin{eqnarray}
{\mathcal G}_{16}^{(1)\delta^2}&=&\frac{1}{2}\delta^{2}|b|^{2}\int\frac{d^{4}kd^{4}\theta_{12}}{(2\pi)^{4}}\theta_{1}^{2}\bar{\theta}_{2}^{2}\left[-\frac{1}{4}\bar{D}_{1}^{2}(k)\langle\Phi_{1}\bar{\Phi}_{1}\rangle\right]\left[-\frac{1}{4}D_{2}^{2}(-k)\langle\bar{\Phi}_{1}\Phi_{1}\rangle\right] \nonumber\\
&=&\frac{\delta^{2}}{2(16)}|b|^{2}\int\frac{d^{4}k}{(2\pi)^{4}}\left\{E(k)E(k){\mathcal J}_{20}(\theta,\bar\theta)+2\frac{|b|^{2}}{16}E(k)B(k){\mathcal J}_{3}(\theta,\bar\theta)\right. \nonumber\\
&&\left.+\frac{|b|^{4}}{(16)^{2}}B(k)B(k){\mathcal J}_{21}(\theta,\bar\theta)\right\} \nonumber\\
&=&\frac{\delta^{2}b^{2}}{2\kappa}\left(\frac{1}{\epsilon}-\overline{\ln}\eta^{2}\right) \nonumber\\
&&+\frac{\delta^{2}b^{2}}{8\kappa b}\left[4b\overline{\ln}\eta^{2}-(\eta^{2}+2b)\overline{\ln}\eta^{+}+(\eta^{2}-2b)\overline{\ln}\eta^{-}+2b\right] \ . \label{G16}
\end{eqnarray}

The sixth set is:
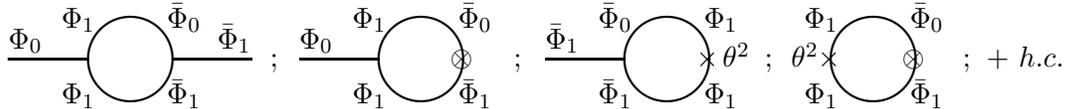
\begin{figure}[htb!]
\begin{eqnarray*}
\begin{picture}(425,5) \thicklines    
\put(15,0){\line(50,0){30}}\put(15,4){$\Phi_0$}\put(35,12){$\Phi_1$}\put(35,-17){$\Phi_1$}\put(61,0){\circle{30}}\put(75,12){$\bar{\Phi}_0$}\put(75,-17){$\bar{\Phi}_1$}\put(77,0){\line(50,0){30}}\put(94,4){$\bar{\Phi}_1$}\put(114,-3){;}
\put(125,0){\line(50,0){30}}\put(125,4){$\Phi_0$}\put(145,12){$\Phi_1$}\put(145,-17){$\Phi_1$}\put(171,0){\circle{30}}\put(185,12){$\bar{\Phi}_0$}\put(185,-17){$\bar{\Phi}_1$}\put(182,-3){$\otimes$}\put(206,-3){;}
\put(218,0){\line(50,0){30}}\put(218,4){$\bar{\Phi}_1$}\put(238,12){$\bar{\Phi}_0$}\put(238,-17){$\bar{\Phi}_1$}\put(264,0){\circle{30}}\put(278,12){$\Phi_1$}\put(278,-17){$\Phi_1$}\put(275,-3){$\times$}\put(285,-3){$\theta^2$}\put(301,-3){;}
\put(311,-3){$\theta^2$}\put(321,-3){$\times$}\put(316,12){$\Phi_1$}\put(316,-17){$\Phi_1$}\put(342,0){\circle{30}}\put(356,12){$\bar{\Phi}_0$}\put(356,-17){$\bar{\Phi}_1$}\put(353,-3){$\otimes$}\put(376,-3){;}
\put(385,-3){+ $h.c.$}
\end{picture}
\end{eqnarray*}
\caption{Diagrams ${\mathcal G}_{17}^{(1)\delta^2}$, ${\mathcal G}_{18}^{(1)\delta^2}$, ${\mathcal G}_{19}^{(1)\delta^2}$ 
and ${\mathcal G}_{20}^{(1)\delta^2}$.}
\label{fig7}
\end{figure}

\begin{eqnarray}
{\mathcal G}_{17}^{(1)\delta^2}&=&16\delta^{2}g^{2}\int\frac{d^{4}kd^{4}\theta_{12}}{(2\pi)^{4}}\Phi_{0}(1)\bar{\Phi}_{1}(2)\left[-\frac{1}{4}\bar{D}_{1}^{2}(k)\langle\Phi_{1}\bar{\Phi}_{1}\rangle\right]\left[-\frac{1}{4}D_{2}^{2}(-k)\langle\bar{\Phi}_{0}\Phi_{1}\rangle\right]+h.c. \nonumber\\
&=&\frac{16\delta^{2}g^{2}}{(16)^{2}}a\bar{b}\int\frac{d^{4}k}{(2\pi)^{4}}C(k)\left\{E(k){\mathcal J}_{22}(\theta,\bar\theta)+\frac{|b|^{2}}{16}B(k){\mathcal J}_{23}(\theta,\bar\theta)\right\}+h.c. \nonumber\\
&=&-\frac{8\delta^{2}g^{2}a\langle F_{0}\rangle\langle\varphi_{1}\rangle}{\kappa}\left(\overline{\ln}\eta^{+}-\overline{\ln}\eta^{-}\right) \ . \label{G17}
\end{eqnarray}
\begin{eqnarray}
{\mathcal G}_{18}^{(1)\delta^2}&=&-4\delta^{2}g\bar{a}\int\frac{d^{4}kd^{4}\theta_{12}}{(2\pi)^{4}}\Phi_{0}(1)\left[-\frac{1}{4}\bar{D}_{1}^{2}(k)\langle\Phi_{1}\bar{\Phi}_{1}\rangle\right]\left[-\frac{1}{4}D_{2}^{2}(-k)\langle\bar{\Phi}_{0}\Phi_{1}\rangle\right]+h.c. \nonumber\\
&=&-\frac{4\delta^{2}g}{(16)^{2}}|a|^{2}\bar{b}\int\frac{d^{4}k}{(2\pi)^{4}}C(k)\left\{E(k){\mathcal J}_{6}(\theta,\bar\theta)+\frac{|b|^{2}}{16}B(k){\mathcal J}_{24}(\theta,\bar\theta)\right\}+h.c. \nonumber\\
&=&\frac{2\delta^{2}ga^{2}\langle F_{0}\rangle}{\kappa}\left(\overline{\ln}\eta^{+}-\overline{\ln}\eta^{-}\right) \ . \label{G18}
\end{eqnarray}
\begin{eqnarray}
{\mathcal G}_{19}^{(1)\delta^2}&=&-4\delta^{2}gb\int\frac{d^{4}kd^{4}\theta_{12}}{(2\pi)^{4}}\bar{\Phi}_{1}(1)\theta_{2}^{2}\left[-\frac{1}{4}\bar{D}_{1}^{2}(k)\langle\bar{\Phi}_{1}\Phi_{1}\rangle\right]\left[-\frac{1}{4}D_{2}^{2}(-k)\langle\Phi_{1}\bar{\Phi}_{0}\rangle\right]+h.c. \nonumber\\
&=&-\frac{4\delta^{2}g}{(16)^{2}}a|b|^{2}\int\frac{d^{4}k}{(2\pi)^{4}}C(k)\left\{E(k){\mathcal J}_{25}(\theta,\bar\theta)+\frac{|b|^{2}}{16}B(k){\mathcal J}_{26}(\theta,\bar\theta)\right\}+h.c. \nonumber\\
&=&\frac{2\delta^{2}gab\langle\varphi_{1}\rangle}{\kappa}\left(\overline{\ln}\eta^{+}-\overline{\ln}\eta^{-}\right) \ . \label{G19}
\end{eqnarray}
\begin{eqnarray}
{\mathcal G}_{20}^{(1)\delta^2}&=&\delta^{2}\bar{a}b\int\frac{d^{4}kd^{4}\theta_{12}}{(2\pi)^{4}}\theta_{1}^{2}\left[-\frac{1}{4}\bar{D}_{1}^{2}(k)\langle\Phi_{1}\bar{\Phi}_{1}\rangle\right]\left[-\frac{1}{4}D_{2}^{2}(-k)\langle\bar{\Phi}_{0}\Phi_{1}\rangle\right]+h.c. \nonumber\\
&=&\frac{\delta^{2}}{(16)^{2}}|a|^{2}|b|^{2}\int\frac{d^{4}k}{(2\pi)^{4}}C(k)\left\{E(k){\mathcal J}_{9}(\theta,\bar\theta)+\frac{|b|^{2}}{16}B(k){\mathcal J}_{27}(\theta,\bar\theta)\right\}+h.c. \nonumber\\
&=&-\frac{\delta^{2}a^{2}b}{2\kappa}\left(\overline{\ln}\eta^{+}-\overline{\ln}\eta^{-}\right) \ . \label{G20}
\end{eqnarray}

The seventh set is:

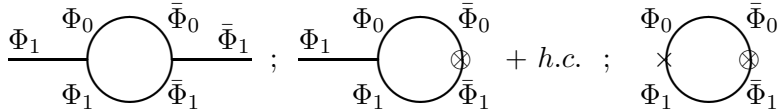
\begin{figure}[htb!]
\begin{eqnarray*}
\begin{picture}(315,5) \thicklines 
\put(15,0){\line(50,0){30}}\put(15,4){$\Phi_1$}\put(35,12){$\Phi_0$}\put(35,-17){$\Phi_1$}\put(61,0){\circle{30}}\put(75,12){$\bar{\Phi}_0$}\put(75,-17){$\bar{\Phi}_1$}\put(77,0){\line(50,0){30}}\put(94,4){$\bar{\Phi}_1$}\put(114,-3){;}
\put(125,0){\line(50,0){30}}\put(125,4){$\Phi_1$}\put(145,12){$\Phi_0$}\put(145,-17){$\Phi_1$}\put(171,0){\circle{30}}\put(185,12){$\bar{\Phi}_0$}\put(185,-17){$\bar{\Phi}_1$}\put(182,-3){$\otimes$}\put(203,-3){+ $h.c.$}\put(240,-3){;}
\put(259,-3){$\times$}\put(254,12){$\Phi_0$}\put(254,-17){$\Phi_1$}\put(280,0){\circle{30}}\put(294,12){$\bar{\Phi}_0$}\put(294,-17){$\bar{\Phi}_1$}\put(291,-3){$\otimes$}
\end{picture}
\end{eqnarray*}
\caption{Diagrams ${\mathcal G}_{21}^{(1)\delta^2}$, ${\mathcal G}_{22}^{(1)\delta^2}$ and ${\mathcal G}_{23}^{(1)\delta^2}$.}
\label{fig8}
\end{figure}

\begin{eqnarray}
{\mathcal G}_{21}^{(1)\delta^2}&=&16\delta^{2}g^{2}\int\frac{d^{4}kd^{4}\theta_{12}}{(2\pi)^{4}}\Phi_{1}(1)\bar{\Phi}_{1}(2)\left[-\frac{1}{4}\bar{D}_{1}^{2}(k)\langle\Phi_{1}\bar{\Phi}_{1}\rangle\right]\left[-\frac{1}{4}D_{2}^{2}(-k)\langle\bar{\Phi}_{0}\Phi_{0}\rangle\right] \nonumber\\
&=&\frac{16\delta^{2}g^{2}}{16}\int\frac{d^{4}k}{(2\pi)^{4}}\left\{(k^{2}+|M|^{2})E(k)A(k){\mathcal J}_{28}(\theta,\bar\theta)+|a|^{2}|b|^{2}E(k)B(k){\mathcal J}_{29}(\theta,\bar\theta)\right. \nonumber\\
&&\left.+(k^{2}+|M|^{2})\frac{|b|^{2}}{16}B(k)A(k){\mathcal J}_{30}(\theta,\bar\theta)+\frac{|a|^{2}|b|^{4}}{16}B(k)B(k){\mathcal J}_{31}(\theta,\bar\theta)\right\} \nonumber\\
&=&\frac{4\delta^{2}g^{2}\langle\varphi_{1}\rangle^{2}}{\kappa b}\left\{4b(2\eta^{2}-M^{2})\overline{\ln}\eta^{2}+[a^{2}\eta^{2}+2\eta^{+}(M^{2}-\eta^{+})]\overline{\ln}\eta^{+}\right. \nonumber\\
&&\left.-[a^{2}\eta^{2}+2\eta^{-}(M^{2}-\eta^{-})]\overline{\ln}\eta^{-}-2b(a^{2}+2M^{2}-2\eta^{2})\right\} \ . \label{G21}
\end{eqnarray}
\begin{eqnarray}
{\mathcal G}_{22}^{(1)\delta^2}&=&-4\delta^{2}g\bar{a}\int\frac{d^{4}kd^{4}\theta_{12}}{(2\pi)^{4}}\Phi_{1}(1)\left[-\frac{1}{4}\bar{D}_{1}^{2}(k)\langle\Phi_{1}\bar{\Phi}_{1}\rangle\right]\left[-\frac{1}{4}D_{2}^{2}(-k)\langle\bar{\Phi}_{0}\Phi_{0}\rangle\right]+h.c. \nonumber\\
&=&-\frac{4\delta^{2}g}{16}\bar{a}\int\frac{d^{4}k}{(2\pi)^{4}}\left\{(k^{2}+|M|^{2})E(k)A(k){\mathcal J}_{13}(\theta,\bar\theta)+|a|^{2}|b|^{2}E(k)B(k){\mathcal J}_{32}(\theta,\bar\theta)\right. \nonumber\\
&&\left.+(k^{2}+|M|^{2})\frac{|b|^{2}}{16}B(k)A(k){\mathcal J}_{7}(\theta,\bar\theta)+\frac{|a|^{2}|b|^{4}}{16}B(k)B(k){\mathcal J}_{8}(\theta,\bar\theta)\right\}+h.c. \nonumber\\
&=&-\frac{2\delta^{2}ga\langle\varphi_{1}\rangle}{\kappa b}\left\{4b(2\eta^{2}-M^{2})\overline{\ln}\eta^{2}+[a^{2}\eta^{2}+2\eta^{+}(M^{2}-\eta^{+})]\overline{\ln}\eta^{+}\right. \nonumber\\
&&\left.-[a^{2}\eta^{2}+2\eta^{-}(M^{2}-\eta^{-})]\overline{\ln}\eta^{-}-2b(a^{2}+2M^{2}-2\eta^{2})\right\} \ . \label{G22}
\end{eqnarray}
\begin{eqnarray}
{\mathcal G}_{23}^{(1)\delta^2}&=&\delta^{2}|a|^{2}\int\frac{d^{4}kd^{4}\theta_{12}}{(2\pi)^{4}}\left[-\frac{1}{4}\bar{D}_{1}^{2}(k)\langle\Phi_{1}\bar{\Phi}_{1}\rangle\right]\left[-\frac{1}{4}D_{2}^{2}(-k)\langle\bar{\Phi}_{0}\Phi_{0}\rangle\right] \nonumber\\
&=&\frac{\delta^{2}}{16}|a|^{2}\int\frac{d^{4}k}{(2\pi)^{4}}\left\{(k^{2}+|M|^{2})E(k)A(k){\mathcal J}_{33}(\theta,\bar\theta)+|a|^{2}|b|^{2}E(k)B(k){\mathcal J}_{34}(\theta,\bar\theta)\right. \nonumber\\
&&\left.+(k^{2}+|M|^{2})\frac{|b|^{2}}{16}B(k)A(k){\mathcal J}_{35}(\theta,\bar\theta)+\frac{|a|^{2}|b|^{4}}{16}B(k)B(k){\mathcal J}_{36}(\theta,\bar\theta)\right\} \nonumber\\
&=&\frac{\delta^{2}a^{2}}{4\kappa b}\left\{4b(2\eta^{2}-M^{2})\overline{\ln}\eta^{2}+[a^{2}\eta^{2}+2\eta^{+}(M^{2}-\eta^{+})]\overline{\ln}\eta^{+}\right. \nonumber\\
&&\left.-[a^{2}\eta^{2}+2\eta^{-}(M^{2}-\eta^{-})]\overline{\ln}\eta^{-}-2b(a^{2}+2M^{2}-2\eta^{2})\right\} \ . \label{G23}
\end{eqnarray}

The eighth set is:

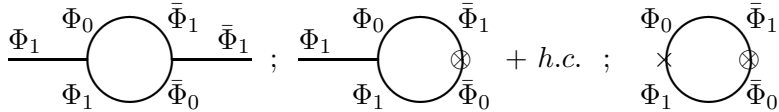
\begin{figure}[htb!]
\begin{eqnarray*}
\begin{picture}(315,5) \thicklines 
\put(15,0){\line(50,0){30}}\put(15,4){$\Phi_1$}\put(35,12){$\Phi_0$}\put(35,-17){$\Phi_1$}\put(61,0){\circle{30}}\put(75,12){$\bar{\Phi}_1$}\put(75,-17){$\bar{\Phi}_0$}\put(77,0){\line(50,0){30}}\put(94,4){$\bar{\Phi}_1$}\put(114,-3){;}
\put(125,0){\line(50,0){30}}\put(125,4){$\Phi_1$}\put(145,12){$\Phi_0$}\put(145,-17){$\Phi_1$}\put(171,0){\circle{30}}\put(185,12){$\bar{\Phi}_1$}\put(185,-17){$\bar{\Phi}_0$}\put(182,-3){$\otimes$}\put(203,-3){+ $h.c.$}\put(240,-3){;}
\put(259,-3){$\times$}\put(254,12){$\Phi_0$}\put(254,-17){$\Phi_1$}\put(280,0){\circle{30}}\put(294,12){$\bar{\Phi}_1$}\put(294,-17){$\bar{\Phi}_0$}\put(291,-3){$\otimes$}
\end{picture}
\end{eqnarray*}
\caption{Diagrams ${\mathcal G}_{24}^{(1)\delta^2}$, ${\mathcal G}_{25}^{(1)\delta^2}$ and ${\mathcal G}_{26}^{(1)\delta^2}$.}
\label{fig9}
\end{figure}

\begin{eqnarray}
{\mathcal G}_{24}^{(1)\delta^2}&=&16\delta^{2}g^{2}\int\frac{d^{4}kd^{4}\theta_{12}}{(2\pi)^{4}}\Phi_{1}(1)\bar{\Phi}_{1}(2)\left[-\frac{1}{4}\bar{D}_{1}^{2}(k)\langle\Phi_{1}\bar{\Phi}_{0}\rangle\right]\left[-\frac{1}{4}D_{2}^{2}(-k)\langle\bar{\Phi}_{1}\Phi_{0}\rangle\right] \nonumber\\
&=&\frac{16\delta^{2}g^{2}}{(16)^{3}}|a|^{2}|b|^{2}\int\frac{d^{4}k}{(2\pi)^{4}}C(k)C(k){\mathcal J}_{37}(\theta,\bar\theta) \nonumber\\
&=&\frac{4\delta^{2}g^{2}a^{2}\langle\varphi_{1}\rangle^{2}}{\kappa b}\left(\eta^{2}\overline{\ln}\eta^{+}-\eta^{2}\overline{\ln}\eta^{-}-2b\right) \ . \label{G24}
\end{eqnarray}
\begin{eqnarray}
{\mathcal G}_{25}^{(1)\delta^2}&=&-4\delta^{2}g\bar{a}\int\frac{d^{4}kd^{4}\theta_{12}}{(2\pi)^{4}}\Phi_{1}(1)\left[-\frac{1}{4}\bar{D}_{1}^{2}(k)\langle\Phi_{1}\bar{\Phi}_{0}\rangle\right]\left[-\frac{1}{4}D_{2}^{2}(-k)\langle\bar{\Phi}_{1}\Phi_{0}\rangle\right]+h.c. \nonumber\\
&=&-\frac{4\delta^{2}g}{(16)^{3}}|a|^{2}\bar{a}|b|^{2}\int\frac{d^{4}k}{(2\pi)^{4}}C(k)C(k){\mathcal J}_{38}(\theta,\bar\theta)+h.c. \nonumber\\
&=&-\frac{2\delta^{2}ga^{3}\langle\varphi_{1}\rangle}{\kappa b}\left(\eta^{2}\overline{\ln}\eta^{+}-\eta^{2}\overline{\ln}\eta^{-}-2b\right) \ . \label{G25}
\end{eqnarray}
\begin{eqnarray}
{\mathcal G}_{26}^{(1)\delta^2}&=&\delta^{2}|a|^{2}\int\frac{d^{4}kd^{4}\theta_{12}}{(2\pi)^{4}}\left[-\frac{1}{4}\bar{D}_{1}^{2}(k)\langle\Phi_{1}\bar{\Phi}_{0}\rangle\right]\left[-\frac{1}{4}D_{2}^{2}(-k)\langle\bar{\Phi}_{1}\Phi_{0}\rangle\right] \nonumber\\
&=&\frac{\delta^{2}}{(16)^{3}}|a|^{4}|b|^{2}\int\frac{d^{4}k}{(2\pi)^{4}}C(k)C(k){\mathcal J}_{39}(\theta,\bar\theta) \nonumber\\
&=&\frac{\delta^{2}a^{4}}{4\kappa b}\left(\eta^{2}\overline{\ln}\eta^{+}-\eta^{2}\overline{\ln}\eta^{-}-2b\right) \ . \label{G26}
\end{eqnarray}

The ninth set is:

\begin{figure}[htb!]
\begin{eqnarray*}
\begin{picture}(440,5) \thicklines 
\put(125,0){\line(50,0){30}}\put(125,4){$\Phi_0$}\put(145,12){$\Phi_1$}\put(145,-17){$\Phi_1$}\put(171,0){\circle{30}}\put(185,12){$\Phi_1$}\put(185,-17){$\Phi_2$}\put(182,-3){$\times$}\put(208,-3){;}
\put(219,-3){$\theta^2$}\put(229,-3){$\times$}\put(224,12){$\Phi_1$}\put(224,-17){$\Phi_1$}\put(250,0){\circle{30}}\put(264,12){$\Phi_1$}\put(264,-17){$\Phi_2$}\put(261,-3){$\times$}\put(286,-3){;}
\put(295,-3){+ $h.c.$}
\end{picture}
\end{eqnarray*}
\caption{Diagrams ${\mathcal G}_{27}^{(1)\delta^2}$ and ${\mathcal G}_{28}^{(1)\delta^2}$.}
\label{fig10}
\end{figure}
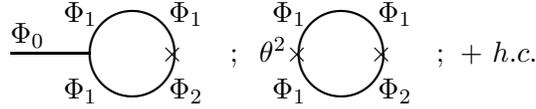

\begin{eqnarray}
{\mathcal G}_{27}^{(1)\delta^2}&=&-4\delta^{2}g\rho\int\frac{d^{4}kd^{4}\theta_{12}}{(2\pi)^{4}}\Phi_{0}(1)\left[-\frac{1}{4}\bar{D}_{1}^{2}(k)\langle\Phi_{1}\Phi_{2}\rangle\right]\left[-\frac{1}{4}\bar{D}_{2}^{2}(-k)\langle\Phi_{1}\Phi_{1}\rangle\right]+h.c. \nonumber\\
&=&-\frac{4\delta^{2}g}{(16)^{2}}\rho\bar{M}\bar{b}\int\frac{d^{4}k}{(2\pi)^{4}}F(k)\left\{A(k){\mathcal J}_{6}(\theta,\bar\theta)-|b|^{2}B(k){\mathcal J}_{2}(\theta,\bar\theta)\right\}+h.c. \nonumber\\
&=&\frac{2\delta^{2}g\rho M\langle F_{0}\rangle}{\kappa}\left(\overline{\ln}\eta^{+}-\overline{\ln}\eta^{-}\right) \ . \label{G27}
\end{eqnarray}
\begin{eqnarray}
{\mathcal G}_{28}^{(1)\delta^2}&=&\delta^{2}b\rho\int\frac{d^{4}kd^{4}\theta_{12}}{(2\pi)^{4}}\theta_{1}^{2}\left[-\frac{1}{4}\bar{D}_{1}^{2}(k)\langle\Phi_{1}\Phi_{2}\rangle\right]\left[-\frac{1}{4}\bar{D}_{2}^{2}(-k)\langle\Phi_{1}\Phi_{1}\rangle\right]+h.c. \nonumber\\
&=&\frac{\delta^{2}}{(16)^{2}}\rho\bar{M}|b|^{2}\int\frac{d^{4}k}{(2\pi)^{4}}F(k)\left\{A(k){\mathcal J}_{9}(\theta,\bar\theta)-|b|^{2}B(k){\mathcal J}_{3}(\theta,\bar\theta)\right\}+h.c. \nonumber\\
&=&-\frac{\delta^{2}\rho Mb}{2\kappa}\left(\overline{\ln}\eta^{+}-\overline{\ln}\eta^{-}\right) \ . \label{G28}
\end{eqnarray}
\newpage
The tenth set is:

\begin{figure}[htb!]
\begin{eqnarray*}
\begin{picture}(440,5) \thicklines 
\put(125,0){\line(50,0){30}}\put(125,4){$\Phi_1$}\put(145,12){$\Phi_0$}\put(145,-17){$\Phi_1$}\put(171,0){\circle{30}}\put(185,12){$\Phi_1$}\put(185,-17){$\Phi_2$}\put(182,-3){$\times$}\put(208,-3){;}
\put(227,-3){$\times$}\put(222,12){$\Phi_0$}\put(222,-17){$\Phi_1$}\put(248,0){\circle{30}}\put(262,12){$\Phi_1$}\put(262,-17){$\Phi_2$}\put(259,-3){$\times$}\put(284,-3){;}
\put(293,-3){+ $h.c.$}
\end{picture}
\end{eqnarray*}
\caption{Diagrams ${\mathcal G}_{29}^{(1)\delta^2}$ and ${\mathcal G}_{30}^{(1)\delta^2}$.}
\label{fig11}
\end{figure}
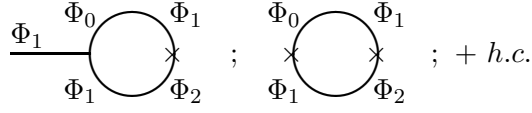

\begin{eqnarray}
{\mathcal G}_{29}^{(1)\delta^2}&=&-4\delta^{2}g\rho\int\frac{d^{4}kd^{4}\theta_{12}}{(2\pi)^{4}}\Phi_{1}(1)\left[-\frac{1}{4}\bar{D}_{1}^{2}(k)\langle\Phi_{1}\Phi_{2}\rangle\right]\left[-\frac{1}{4}\bar{D}_{2}^{2}(-k)\langle\Phi_{1}\Phi_{0}\rangle\right]+h.c. \nonumber\\
&=&-\frac{4\delta^{2}g}{(16)^{2}}\rho\bar{M}\bar{a}\int\frac{d^{4}k}{(2\pi)^{4}}\left\{A(k)A(k){\mathcal J}_{13}(\theta,\bar\theta)-2|b|^{2}A(k)B(k){\mathcal J}_{7}(\theta,\bar\theta)\right. \nonumber\\
&&\left.+|b|^{4}B(k)B(k){\mathcal J}_{8}(\theta,\bar\theta)\right\}+h.c. \nonumber\\
&=&-\frac{2\delta^{2}g\rho Ma\langle\varphi_{1}\rangle}{\kappa b}\left[4b\overline{\ln}\eta^{2}-(\eta^{2}+2b)\overline{\ln}\eta^{+}+(\eta^{2}-2b)\overline{\ln}\eta^{-}+2b\right] \ . \label{G29}
\end{eqnarray}
\begin{eqnarray}
{\mathcal G}_{30}^{(1)\delta^2}&=&\delta^{2}a\rho\int\frac{d^{4}kd^{4}\theta_{12}}{(2\pi)^{4}}\left[-\frac{1}{4}\bar{D}_{1}^{2}(k)\langle\Phi_{1}\Phi_{2}\rangle\right]\left[-\frac{1}{4}\bar{D}_{2}^{2}(-k)\langle\Phi_{1}\Phi_{0}\rangle\right]+h.c. \nonumber\\
&=&\frac{\delta^{2}}{(16)^{2}}\rho\bar{M}|a|^{2}\int\frac{d^{4}k}{(2\pi)^{4}}\left\{A(k)A(k){\mathcal J}_{14}(\theta,\bar\theta)-2|b|^{2}A(k)B(k){\mathcal J}_{9}(\theta,\bar\theta)\right. \nonumber\\
&&\left.+|b|^{4}B(k)B(k){\mathcal J}_{36}(\theta,\bar\theta)\right\}+h.c. \nonumber\\
&=&\frac{\delta^{2}\rho Ma^{2}}{2\kappa b}\left[4b\overline{\ln}\eta^{2}-(\eta^{2}+2b)\overline{\ln}\eta^{+}+(\eta^{2}-2b)\overline{\ln}\eta^{-}+2b\right] \ . \label{G30}
\end{eqnarray}

The eleventh set is:

\begin{figure}[htb!]
\begin{eqnarray*}
\begin{picture}(440,5) \thicklines 
\put(125,0){\line(50,0){30}}\put(125,4){$\Phi_1$}\put(145,12){$\Phi_0$}\put(145,-17){$\Phi_1$}\put(171,0){\circle{30}}\put(185,12){$\Phi_2$}\put(185,-17){$\Phi_1$}\put(182,-3){$\times$}\put(208,-3){;}
\put(227,-3){$\times$}\put(222,12){$\Phi_0$}\put(222,-17){$\Phi_1$}\put(248,0){\circle{30}}\put(262,12){$\Phi_2$}\put(262,-17){$\Phi_1$}\put(259,-3){$\times$}\put(284,-3){;}
\put(293,-3){+ $h.c.$}
\end{picture}
\end{eqnarray*}
\caption{Diagrams ${\mathcal G}_{31}^{(1)\delta^2}$ and ${\mathcal G}_{32}^{(1)\delta^2}$.}
\label{fig12}
\end{figure}
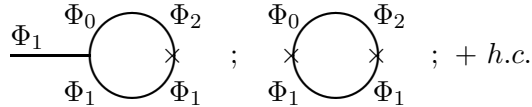

\begin{eqnarray}
{\mathcal G}_{31}^{(1)\delta^2}&=&-4\delta^{2}g\rho\int\frac{d^{4}kd^{4}\theta_{12}}{(2\pi)^{4}}\Phi_{1}(1)\left[-\frac{1}{4}\bar{D}_{1}^{2}(k)\langle\Phi_{1}\bar{\Phi}_{1}\rangle\right]\left[-\frac{1}{4}\bar{D}_{2}^{2}(-k)\langle\Phi_{2}\Phi_{0}\rangle\right]+h.c. \nonumber\\
&=&\frac{4\delta^{2}g}{(16)^{2}}\rho\bar{M}\bar{a}\bar{b}^{2}\int\frac{d^{4}k}{(2\pi)^{4}}F(k)C(k){\mathcal J}_{7}(\theta,\bar\theta)+h.c. \nonumber\\
&=&-\frac{2\delta^{2}g\rho Ma\langle\varphi_{1}\rangle}{\kappa b}\left(\eta^{2}\overline{\ln}\eta^{+}-\eta^{2}\overline{\ln}\eta^{-}-2b\right) \ . \label{G31}
\end{eqnarray}
\begin{eqnarray}
{\mathcal G}_{32}^{(1)\delta^2}&=&\delta^{2}a\rho\int\frac{d^{4}kd^{4}\theta_{12}}{(2\pi)^{4}}\left[-\frac{1}{4}\bar{D}_{1}^{2}(k)\langle\Phi_{1}\bar{\Phi}_{1}\rangle\right]\left[-\frac{1}{4}\bar{D}_{2}^{2}(-k)\langle\Phi_{2}\Phi_{0}\rangle\right]+h.c. \nonumber\\
&=&-\frac{\delta^{2}}{(16)^{2}}\rho\bar{M}|a|^{2}\bar{b}^{2}\int\frac{d^{4}k}{(2\pi)^{4}}F(k)C(k){\mathcal J}_{9}(\theta,\bar\theta)+h.c. \nonumber\\
&=&\frac{\delta^{2}\rho Ma^{2}}{2\kappa b}\left(\eta^{2}\overline{\ln}\eta^{+}-\eta^{2}\overline{\ln}\eta^{-}-2b\right) \ . \label{G32}
\end{eqnarray}

The twelfth set is:

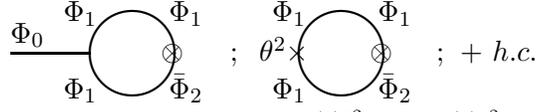
\begin{figure}[htb!]
\begin{eqnarray*}
\begin{picture}(440,5) \thicklines 
\put(125,0){\line(50,0){30}}\put(125,4){$\Phi_0$}\put(145,12){$\Phi_1$}\put(145,-17){$\Phi_1$}\put(171,0){\circle{30}}\put(185,12){$\bar{\Phi}_1$}\put(185,-17){$\bar{\Phi}_2$}\put(182,-3){$\otimes$}\put(208,-3){;}
\put(219,-3){$\theta^2$}\put(229,-3){$\times$}\put(224,12){$\Phi_1$}\put(224,-17){$\Phi_1$}\put(250,0){\circle{30}}\put(264,12){$\bar{\Phi}_1$}\put(264,-17){$\bar{\Phi}_2$}\put(261,-3){$\otimes$}\put(286,-3){;}
\put(295,-3){+ $h.c.$}
\end{picture}
\end{eqnarray*}
\caption{Diagrams ${\mathcal G}_{33}^{(1)\delta^2}$ and ${\mathcal G}_{34}^{(1)\delta^2}$.}
\label{fig13}
\end{figure}

\begin{eqnarray}
{\mathcal G}_{33}^{(1)\delta^2}&=&-4\delta^{2}g\bar{\rho}\int\frac{d^{4}kd^{4}\theta_{12}}{(2\pi)^{4}}\Phi_{0}(1)\left[-\frac{1}{4}\bar{D}_{1}^{2}(k)\langle\Phi_{1}\bar{\Phi}_{2}\rangle\right]\left[-\frac{1}{4}D_{2}^{2}(-k)\langle\bar{\Phi}_{1}\Phi_{1}\rangle\right]+h.c. \nonumber\\
&=&\frac{4\delta^{2}g}{16}\bar{\rho}M\bar{b}\int\frac{d^{4}k}{(2\pi)^{4}}F(k)\left\{E(k){\mathcal J}_{18}(\theta,\bar\theta)+\frac{|b|^{2}}{16}B(k){\mathcal J}_{2}(\theta,\bar\theta)\right\}+h.c. \nonumber\\
&=&\frac{2\delta^{2}g\rho M\langle F_{0}\rangle}{\kappa}\left(\overline{\ln}\eta^{+}-\overline{\ln}\eta^{-}\right) \ . \label{G33}
\end{eqnarray}
\begin{eqnarray}
{\mathcal G}_{34}^{(1)\delta^2}&=&\delta^{2}b\bar{\rho}\int\frac{d^{4}kd^{4}\theta_{12}}{(2\pi)^{4}}\theta_{1}^{2}\left[-\frac{1}{4}\bar{D}_{1}^{2}(k)\langle\Phi_{1}\bar{\Phi}_{2}\rangle\right]\left[-\frac{1}{4}D_{2}^{2}(-k)\langle\bar{\Phi}_{1}\Phi_{1}\rangle\right]+h.c. \nonumber\\
&=&-\frac{\delta^{2}}{16}\bar{\rho}M|b|^{2}\int\frac{d^{4}k}{(2\pi)^{4}}F(k)\left\{E(k){\mathcal J}_{20}(\theta,\bar\theta)+\frac{|b|^{2}}{16}B(k){\mathcal J}_{36}(\theta,\bar\theta)\right\}+h.c. \nonumber\\
&=&-\frac{\delta^{2}\rho Mb}{2\kappa}\left(\overline{\ln}\eta^{+}-\overline{\ln}\eta^{-}\right) \ . \label{G34}
\end{eqnarray}

\vspace{1cm}
The thirteenth set is:

\begin{figure}[htb!]
\begin{eqnarray*}
\begin{picture}(440,5) \thicklines 
\put(125,0){\line(50,0){30}}\put(125,4){$\Phi_1$}\put(145,12){$\Phi_0$}\put(145,-17){$\Phi_1$}\put(171,0){\circle{30}}\put(185,12){$\bar{\Phi}_1$}\put(185,-17){$\bar{\Phi}_2$}\put(182,-3){$\otimes$}\put(208,-3){;}
\put(227,-3){$\times$}\put(222,12){$\Phi_0$}\put(222,-17){$\Phi_1$}\put(248,0){\circle{30}}\put(262,12){$\bar{\Phi}_1$}\put(262,-17){$\bar{\Phi}_2$}\put(259,-3){$\otimes$}\put(284,-3){;}
\put(293,-3){+ $h.c.$}
\end{picture}
\end{eqnarray*}
\caption{Diagrams ${\mathcal G}_{35}^{(1)\delta^2}$ and ${\mathcal G}_{36}^{(1)\delta^2}$.}
\label{fig14}
\end{figure}
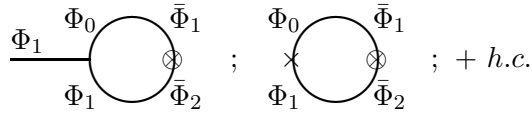

\begin{eqnarray}
{\mathcal G}_{35}^{(1)\delta^2}&=&-4\delta^{2}g\bar{\rho}\int\frac{d^{4}kd^{4}\theta_{12}}{(2\pi)^{4}}\Phi_{1}(1)\left[-\frac{1}{4}\bar{D}_{1}^{2}(k)\langle\Phi_{1}\bar{\Phi}_{2}\rangle\right]\left[-\frac{1}{4}D_{2}^{2}(-k)\langle\bar{\Phi}_{1}\Phi_{0}\rangle\right]+h.c. \nonumber\\
&=&\frac{4\delta^{2}g}{(16)^{2}}\bar{\rho}M\bar{a}|b|^{2}\int\frac{d^{4}k}{(2\pi)^{4}}F(k)C(k){\mathcal J}_{7}(\theta,\bar\theta)+h.c. \nonumber\\
&=&-\frac{2\delta^{2}g\rho Ma\langle\varphi_{1}\rangle}{\kappa b}\left(\eta^{2}\overline{\ln}\eta^{+}-\eta^{2}\overline{\ln}\eta^{-}-2b\right) \ . \label{G35}
\end{eqnarray}
\begin{eqnarray}
{\mathcal G}_{36}^{(1)\delta^2}&=&\delta^{2}a\bar{\rho}\int\frac{d^{4}kd^{4}\theta_{12}}{(2\pi)^{4}}\left[-\frac{1}{4}\bar{D}_{1}^{2}(k)\langle\Phi_{1}\bar{\Phi}_{2}\rangle\right]\left[-\frac{1}{4}D_{2}^{2}(-k)\langle\bar{\Phi}_{1}\Phi_{0}\rangle\right]+h.c. \nonumber\\
&=&-\frac{\delta^{2}}{(16)^{2}}\bar{\rho}M|a|^{2}|b|^{2}\int\frac{d^{4}k}{(2\pi)^{4}}F(k)C(k){\mathcal J}_{9}(\theta,\bar\theta)+h.c. \nonumber\\
&=&\frac{\delta^{2}\rho Ma^{2}}{2\kappa b}\left(\eta^{2}\overline{\ln}\eta^{+}-\eta^{2}\overline{\ln}\eta^{-}-2b\right) \ . \label{G36}
\end{eqnarray}

The fourteenth set is:

\begin{figure}[htb!]
\begin{eqnarray*}
\begin{picture}(440,5) \thicklines 
\put(125,0){\line(50,0){30}}\put(125,4){$\Phi_1$}\put(145,12){$\Phi_0$}\put(145,-17){$\Phi_1$}\put(171,0){\circle{30}}\put(185,12){$\bar{\Phi}_2$}\put(185,-17){$\bar{\Phi}_1$}\put(182,-3){$\otimes$}\put(208,-3){;}
\put(227,-3){$\times$}\put(222,12){$\Phi_0$}\put(222,-17){$\Phi_1$}\put(248,0){\circle{30}}\put(262,12){$\bar{\Phi}_2$}\put(262,-17){$\bar{\Phi}_1$}\put(259,-3){$\otimes$}\put(284,-3){;}
\put(293,-3){+ $h.c.$}
\end{picture}
\end{eqnarray*}
\caption{Diagrams ${\mathcal G}_{37}^{(1)\delta^2}$ and ${\mathcal G}_{38}^{(1)\delta^2}$.}
\label{fig15}
\end{figure}
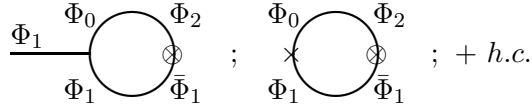

\begin{eqnarray}
{\mathcal G}_{37}^{(1)\delta^2}&=&-4\delta^{2}g\bar{\rho}\int\frac{d^{4}kd^{4}\theta_{12}}{(2\pi)^{4}}\Phi_{1}(1)\left[-\frac{1}{4}\bar{D}_{1}^{2}(k)\langle\Phi_{1}\bar{\Phi}_{1}\rangle\right]\left[-\frac{1}{4}D_{2}^{2}(-k)\langle\bar{\Phi}_{2}\Phi_{0}\rangle\right]+h.c. \nonumber\\
&=&-\frac{4\delta^{2}g}{16}\bar{\rho}M\bar{a}\int\frac{d^{4}k}{(2\pi)^{4}}\left\{-E(k)A(k){\mathcal J}_{40}(\theta,\bar\theta)+|b|^{2}E(k)B(k){\mathcal J}_{32}(\theta,\bar\theta)\right. \nonumber\\
&&\left.-\frac{|b|^{2}}{16}B(k)A(k){\mathcal J}_{7}(\theta,\bar\theta)+\frac{|b|^{4}}{16}B(k)B(k){\mathcal J}_{8}(\theta,\bar\theta)\right\}+h.c. \nonumber\\
&=&-\frac{2\delta^{2}g\rho Ma\langle\varphi_{1}\rangle}{\kappa b}\left[4b\overline{\ln}\eta^{2}-(\eta^{2}+2b)\overline{\ln}\eta^{+}+(\eta^{2}-2b)\overline{\ln}\eta^{-}+2b\right] \ . \label{G37}
\end{eqnarray}
\begin{eqnarray}
{\mathcal G}_{38}^{(1)\delta^2}&=&\delta^{2}a\bar{\rho}\int\frac{d^{4}kd^{4}\theta_{12}}{(2\pi)^{4}}\left[-\frac{1}{4}\bar{D}_{1}^{2}(k)\langle\Phi_{1}\bar{\Phi}_{1}\rangle\right]\left[-\frac{1}{4}D_{2}^{2}(-k)\langle\bar{\Phi}_{2}\Phi_{0}\rangle\right]+h.c. \nonumber\\
&=&\frac{\delta^{2}}{16}\bar{\rho}M|a|^{2}\int\frac{d^{4}k}{(2\pi)^{4}}\left\{-E(k)A(k){\mathcal J}_{33}(\theta,\bar\theta)+|b|^{2}E(k)B(k){\mathcal J}_{34}(\theta,\bar\theta)\right. \nonumber\\
&&\left.-\frac{|b|^{2}}{16}B(k)A(k){\mathcal J}_{35}(\theta,\bar\theta)+\frac{|b|^{4}}{16}B(k)B(k){\mathcal J}_{36}(\theta,\bar\theta)\right\}+h.c. \nonumber\\
&=&\frac{\delta^{2}\rho Ma^{2}}{2\kappa b}\left[4b\overline{\ln}\eta^{2}-(\eta^{2}+2b)\overline{\ln}\eta^{+}+(\eta^{2}-2b)\overline{\ln}\eta^{-}+2b\right] \ . \label{G38}
\end{eqnarray}

The last four diagrams have unique propagator structure in the loop and are not divided into sets.

The thirtieth ninth diagram is:

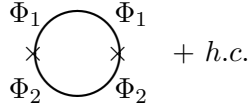
\begin{figure}[htb!]
\begin{eqnarray*}
\begin{picture}(435,5) \thicklines 
\put(182,-3){$\times$}\put(177,12){$\Phi_1$}\put(177,-17){$\Phi_2$}\put(203,0){\circle{30}}\put(217,12)
{$\Phi_1$}\put(217,-17){$\Phi_2$}\put(214,-3){$\times$}\put(239,-3){+ $h.c.$}
\end{picture}
\end{eqnarray*}
\caption{Diagram ${\mathcal G}_{39}^{(1)\delta^2}$.}
\label{fig16}
\end{figure}

\begin{eqnarray}
{\mathcal G}_{39}^{(1)\delta^2}&=&\frac{1}{2}\delta^{2}\rho^{2}\int\frac{d^{4}kd^{4}\theta_{12}}{(2\pi)^{4}}\left[-\frac{1}{4}\bar{D}_{1}^{2}(k)\langle\Phi_{2}\Phi_{2}\rangle\right]\left[-\frac{1}{4}\bar{D}_{2}^{2}(-k)\langle\Phi_{1}\Phi_{1}\rangle\right]+h.c. \nonumber\\
&=&-\frac{\delta^{2}}{2(16)^{2}}\rho^{2}|M|^{2}\bar{b}^{2}\int\frac{d^{4}k}{(2\pi)^{4}}C(k)F(k){\mathcal J}_{9}(\theta,\bar\theta)+h.c. \nonumber\\
&=&\frac{\delta^{2}\rho^{2}M^{2}}{4\kappa b}\left(\eta^{2}\overline{\ln}\eta^{+}-\eta^{2}\overline{\ln}\eta^{-}-2b\right) \ . \label{G39}
\end{eqnarray}

The fortieth diagram is:

\begin{figure}[htb!]
\begin{eqnarray*}
\begin{picture}(392,5) \thicklines 
\put(182,-3){$\times$}\put(177,12){$\Phi_1$}\put(177,-17){$\Phi_2$}\put(203,0){\circle{30}}\put(217,12)
{$\bar{\Phi}_1$}\put(217,-17){$\bar{\Phi}_2$}\put(214,-3){$\times$}
\end{picture}
\end{eqnarray*}
\caption{Diagram ${\mathcal G}_{40}^{(1)\delta^2}$. }
\label{fig17}
\end{figure}
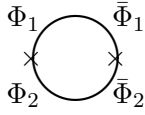

\begin{eqnarray}
{\mathcal G}_{40}^{(1)\delta^2}&=&\delta^{2}|\rho|^{2}\int\frac{d^{4}kd^{4}\theta_{12}}{(2\pi)^{4}}\left[-\frac{1}{4}\bar{D}_{1}^{2}(k)\langle\Phi_{2}\bar{\Phi}_{2}\rangle\right]\left[-\frac{1}{4}D_{2}^{2}(-k)\langle\bar{\Phi}_{1}\Phi_{1}\rangle\right] \nonumber\\
&=&\frac{\delta^{2}}{16}|\rho|^{2}\int\frac{d^{4}k}{(2\pi)^{4}}\left\{(k^{2}+|a|^{2})A(k)E(k){\mathcal J}_{33}(\theta,\bar\theta)+(k^{2}+|a|^{2})\frac{|b|^{2}}{16}A(k)B(k){\mathcal J}_{35}(\theta,\bar\theta)\right. \nonumber\\
&&\left.+|M|^{2}|b|^{2}B(k)E(k){\mathcal J}_{34}(\theta,\bar\theta)+\frac{|M|^{2}|b|^{4}}{16}B(k)B(k){\mathcal J}_{36}(\theta,\bar\theta)\right\} \nonumber\\
&=&\frac{\delta^{2}\rho^{2}}{4\kappa b}\left\{4b(2\eta^{2}-a^{2})\overline{\ln}\eta^{2}-[2\eta^{+}(\eta^{+}-a^{2})-M^{2}\eta^{2}]\overline{\ln}\eta^{+}\right. \nonumber\\
&&\left.+[2\eta^{-}(\eta^{-}-a^{2})-M^{2}\eta^{2}]\overline{\ln}\eta^{-}+2b(2\eta^{2}-2a^{2}-M^{2})\right\} \ . \label{G40}
\end{eqnarray}

The fortieth first diagram is:

\begin{figure}[htb!]
\begin{eqnarray*}
\begin{picture}(435,5) \thicklines 
\put(182,-3){$\times$}\put(177,12){$\Phi_1$}\put(177,-17){$\Phi_2$}\put(203,0){\circle{30}}\put(217,12){$\Phi_2$}\put(217,-17){$\Phi_1$}\put(214,-3){$\times$}\put(239,-3){+ $h.c.$}
\end{picture}
\end{eqnarray*}
\caption{Diagram ${\mathcal G}_{41}^{(1)\delta^2}$.}
\label{fig18}
\end{figure}
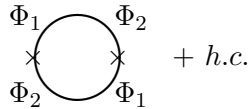

\begin{eqnarray}
{\mathcal G}_{41}^{(1)\delta^2}&=&\frac{1}{2}\delta^{2}\rho^{2}\int\frac{d^{4}kd^{4}\theta_{12}}{(2\pi)^{4}}\left[-\frac{1}{4}\bar{D}_{1}^{2}(k)\langle\Phi_{2}\Phi_{1}\rangle\right]\left[-\frac{1}{4}\bar{D}_{2}^{2}(-k)\langle\Phi_{2}\Phi_{1}\rangle\right]+h.c. \nonumber\\
&=&\frac{\delta^{2}}{2(16)^{2}}\rho^{2}\bar{M}^{2}\int\frac{d^{4}k}{(2\pi)^{4}}\left\{A(k)A(k){\mathcal J}_{14}(\theta,\bar\theta)-2|b|^{2}A(k)B(k){\mathcal J}_{9}(\theta,\bar\theta)\right. \nonumber\\
&&\left.+|b|^{4}B(k)B(k){\mathcal J}_{3}(\theta,\bar\theta)\right\}+h.c. \nonumber\\
&=&\frac{\delta^{2}\rho^{2}M^{2}}{4\kappa b}\left[4b\overline{\ln}\eta^{2}-(\eta^{2}+2b)\overline{\ln}\eta^{+}+(\eta^{2}-2b)\overline{\ln}\eta^{-}+2b\right] \ . \label{G41}
\end{eqnarray}

Finally, the fortieth second diagram is:

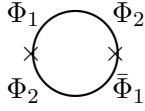
\begin{figure}[htb!]
\begin{eqnarray*}
\begin{picture}(392,5) \thicklines 
\put(182,-3){$\times$}\put(177,12){$\Phi_1$}\put(177,-17){$\Phi_2$}\put(203,0){\circle{30}}\put(217,12){$\bar{\Phi}_2$}\put(217,-17){$\bar{\Phi}_1$}\put(214,-3){$\times$}
\end{picture}
\end{eqnarray*}
\caption{Diagram ${\mathcal G}_{42}^{(1)\delta^2}$.}
\label{fig19}
\end{figure}

\begin{eqnarray}
{\mathcal G}_{42}^{(1)\delta^2}&=&\delta^{2}|\rho|^{2}\int\frac{d^{4}kd^{4}\theta_{12}}{(2\pi)^{4}}\left[-\frac{1}{4}\bar{D}_{1}^{2}(k)
\langle\Phi_{2}\bar{\Phi}_{1}\rangle\right]\left[-\frac{1}{4}D_{2}^{2}(-k)\langle\bar{\Phi}_{2}\Phi_{1}\rangle\right] \nonumber\\
&=&\frac{\delta^{2}}{16}|\rho|^{2}|M|^{2}|b|^{2}\int\frac{d^{4}k}{(2\pi)^{4}}F(k)F(k){\mathcal J}_{20}(\theta,\bar\theta) \nonumber\\
&=&\frac{\delta^{2}\rho^{2}M^{2}}{4\kappa b}\left(\eta^{2}\overline{\ln}\eta^{+}-\eta^{2}\overline{\ln}\eta^{-}-2b\right) \ . \label{G42}
\end{eqnarray}

The diagrams ${\mathcal G}_{14}^{(1)\delta^2}$, ${\mathcal G}_{15}^{(1)\delta^2}$ and ${\mathcal G}_{16}^{(1)\delta^2}$ of the fifth set give 
divergent contributions and we will discuss their renormalization later.

\subsection{Two Loops}
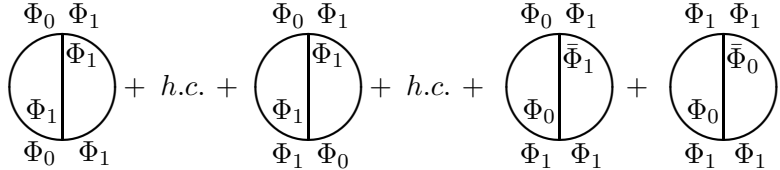
\begin{figure}[htb!]
\begin{eqnarray*}
\begin{picture}(270,10) \thicklines 
\put(15,0){\circle{40}}\put(15,-20){\line(0,50){40}}\put(0,24){$\Phi_0$}\put(0,-28){$\Phi_0$}\put(17,24){$\Phi_1$}\put(21,-28){$\Phi_1$}\put(1,-12)
{$\Phi_1$}\put(16,9){$\Phi_1$}\put(38,-3){+}\put(52,-3){$h.c.$}\put(73,-3){+}
\put(108,0){\circle{40}}\put(108,-20){\line(0,50){40}}\put(94,24){$\Phi_0$}\put(94,-29){$\Phi_1$}\put(111,24){$\Phi_1$}\put(111,-29){$\Phi_0$}
\put(94,-12){$\Phi_1$}\put(109,9){$\Phi_1$}\put(131,-3){+}\put(145,-3){$h.c.$}\put(166,-3){+}
\put(203,0){\circle{40}}\put(203,-20){\line(0,50){40}}\put(188,24){$\bar{\Phi}_0$}\put(188,-29){$\Phi_1$}\put(205,24){$\bar{\Phi}_1$}\put(207,-29)
{$\Phi_1$}\put(189,-12){$\Phi_0$}\put(204,7){$\bar{\Phi}_1$}\put(228,-3){+}
\put(265,0){\circle{40}}\put(265,-20){\line(0,50){40}}\put(250,24){$\bar{\Phi}_1$}\put(250,-29){$\Phi_1$}\put(267,24){$\bar{\Phi}_1$}\put(269,-29)
{$\Phi_1$}\put(251,-12){$\Phi_0$}\put(266,7){$\bar{\Phi}_0$}
\end{picture}
\end{eqnarray*}
\vspace{0.1cm}
\caption{Diagrams ${\mathcal G}_{1}^{(2)\delta^2}$, ${\mathcal G}_{2}^{(2)\delta^2}$, ${\mathcal G}_{3}^{(2)\delta^2}$ 
and ${\mathcal G}_{4}^{(2)\delta^2}$.}
\label{fig20}
\end{figure}

The two-loop diagrams are shown in  Fig. \ref{fig20}.

In order to calculated the two-loop diagrams we are going to use technique developed in \cite{background}. 
The contribution of the first diagram is given by
\begin{eqnarray}
{\mathcal G}_{1}^{(2)\delta^{2}}&=&4\delta^{2}g^{2}\!\!\int\!\frac{d^{4}pd^{4}kd^{4}\theta_{12}}{(2\pi)^8}\!\left[\!-\frac{1}{4}\bar{D}_{1}^{2}(p)\langle\Phi_{0}\Phi_{0}\rangle\right]\!\left[\!-\frac{1}{4}\bar{D}_{2}^{2}(k)\langle\Phi_{1}\Phi_{1}\rangle\right]\!\left[\!\frac{1}{16}\bar{D}_{1}^{2}(q)\bar{D}_{2}^{2}(-q)\langle\Phi_{1}\Phi_{1}\rangle\right] \nonumber\\
&=&-\frac{\delta^{2}g^{2}}{(16)^{3}}a^{2}b^{3}\int\!\frac{d^{4}pd^{4}k}{(2\pi)^8}C(p)F(k)F(q){\mathcal I}_{1}(\theta,\bar\theta) \ , \label{I1}
\end{eqnarray}
with $q=(k-p)$ and the integral ${\mathcal I}_{1}(\theta,\bar\theta)$ is given in Appendix A. Plugging 
${\mathcal I}_{1}(\theta,\bar\theta)$ into (\ref{I1}), we obtain:    
\begin{equation}
{\mathcal G}_{1}^{(2)\delta^{2}}=-4\delta^{2}g^{2}a^{2}b^{3}\int\!\frac{d^{4}pd^{4}k}
{(2\pi)^8}C(p)F(k)F(q)p^{2} \ . \label{diagram 1}
\end{equation}

To handle with the integral above and the other momentum space integrals that will appear in the following, 
we have adopted the strategy: for each of them, we split the integrand with the help of the method of partial 
fraction decomposition and write each integral as the sum of other integrals with just three terms in the 
denominator. The remaining integrals are well known in the literature, and we use the results of the references 
\cite{Jones,Espinosa,Martin} to compute them. 

From now on, we define $\eta^{2}=m^{2}+a^{2}$, $\eta^{\pm}=m^{2}+a^{2}\pm b$ and adopt the 
same notation of references \cite{Jones,Espinosa,Martin} for the integrals $I(x,y,z)$ , $J(x,y)$ and $J(x)$:
\begin{equation}
\kappa J(x)=-\frac{x}{\epsilon}+x\left(\bar{\ln}x-1\right) \ , \label{J(x)}
\end{equation}
\begin{equation}
\kappa^{2}J(x,y)=xy\left[\frac{1}{\epsilon^{2}}+\frac{1}{\epsilon}\left(2-\bar{\ln}x-\bar{\ln}y\right)+\left(1-\bar{\ln}x-\bar{\ln}y+\bar{\ln}x\bar{\ln}y\right)\right] \ , \label{J(x,y)}
\end{equation}
\begin{eqnarray}
\kappa^{2}I(x,y,z)&=&-\frac{c}{2\epsilon^2}-\frac{1}{\epsilon}\left(\frac{3c}{2}-L_1\right)-\frac{1}{2}\left\{L_2-6L_1+(y\!+\!z\!-\!x)\bar{\ln}y\bar{\ln}z\right. \nonumber\\
&&\left.+(z\!+\!x\!-\!y)\bar{\ln}z\bar{\ln}x+(y\!+\!x\!-\!z)\bar{\ln}y\bar{\ln}x+\xi(x,y,z)+c\left[7+\zeta(2)\right]\right\} \ , \label{I(x,y,z)}
\end{eqnarray}
where
\begin{eqnarray*}
\kappa&=&(4\pi)^{2} \ , \\
c&=&x+y+z \ , \\
\bar{\ln}X&=&\ln\left(\frac{X}{\mu^{2}}\right)+\gamma-\ln4\pi \ , \\
L_m&=&x\bar{\ln}^{m}x+y\bar{\ln}^{m}y+z\bar{\ln}^{m}z \ , \\
\xi(x,y,z)&=&S\left[2\ln\frac{z+x-y-S}{2z}\ln\frac{z+y-x-S}{2z}-\ln\frac{x}{z}\ln\frac{y}{z}\right. \\
&&\left. \ \ \ \ \ -2\mbox{Li}_{2}\left(\frac{z+x-y-S}{2z}\right)-2\mbox{Li}_{2}\left(\frac{z+y-x-S}{2z}\right)+\frac{\pi^2}{3}\right] \ , \\
S&=&\sqrt{x^2+y^2+z^2-2xy-2yz-2zx} \ , \\
\mbox{Li}_{2}(z)&=&-\int_{0}^{z}\frac{\ln(1-t)}{t}dt \ \ (\mbox{dilogarithm function}) \ .
\end{eqnarray*}

So, the total contribution of the first two-loop diagram is:
\begin{equation}
{\mathcal G}_{1}^{(2)\delta^{2}}=\frac{\delta^{2}g^{2}a^{2}}{2}\left[I(\eta^{+},\eta^{+},\eta^{+})\!-\!3I(\eta^{+},\eta^{+},\eta^{-})\!+\!3I(\eta^{+},\eta^{-},\eta^{-})\!-\!I(\eta^{-},\eta^{-},\eta^{-})\right] \ , \label{diagram 1 integrals}
\end{equation}

The contribution of the second diagram is given by
\begin{eqnarray}
{\mathcal G}_{2}^{(2)\delta^{2}}&=&8\delta^{2}g^{2}\!\!\int\!\frac{d^{4}pd^{4}kd^{4}\theta_{12}}{(2\pi)^8}\!\!\left[\!-\frac{1}{4}\bar{D}_{1}^{2}(p)\langle\Phi_{0}\Phi_{1}\rangle\right]\!\!\left[\!-\frac{1}{4}\bar{D}_{2}^{2}(k)\langle\Phi_{0}\Phi_{1}\rangle\right]\!\!\left[\!\frac{1}{16}\bar{D}_{1}^{2}(q)\bar{D}_{2}^{2}(-q)\langle\Phi_{1}\Phi_{1}\rangle\right] \nonumber\\
&=&\frac{2\delta^{2}g^{2}}{(16)^{3}}a^{2}b\int\!\frac{d^{4}pd^{4}k}{(2\pi)^8}\left\{A(p)A(k)F(q){\mathcal I}_{2}(\theta,\bar\theta)-2b^{2}A(p)B(k)F(q){\mathcal I}_{3}(\theta,\bar\theta)\right. \nonumber\\
&& \left.+b^{4}B(p)B(k)F(q){\mathcal I}_{4}(\theta,\bar\theta)\right\} \ . \label{I2}
\end{eqnarray}
Plugging ${\mathcal I}_{2}(\theta,\bar\theta)$, ${\mathcal I}_{3}(\theta,\bar\theta)$ and ${\mathcal I}_{4}
(\theta,\bar\theta)$ into (\ref{I2}), we obtain:    
\begin{equation}
{\mathcal G}_{2}^{(2)\delta^{2}}=-8\delta^{2}g^{2}a^{2}b^{3}\left\{2\int\!\frac{d^{4}pd^{4}k}{(2\pi)^8}A(p)B(k)F(q)p^{2}+b^{2}\int\!\frac{d^{4}pd^{4}k}{(2\pi)^8}B(p)B(k)F(q)\right\} \ . \label{second diagram}
\end{equation}
Decomposing the momentum space integrals we get the contribution:
\begin{eqnarray}
{\mathcal G}_{2}^{(2)\delta^{2}}&=&\delta^{2}g^{2}a^{2}\left[-4I(\eta^{2},\eta^{2},\eta^{+})+4I(\eta^{2},\eta^{2},\eta^{-})+I(\eta^{+},\eta^{+},\eta^{+})\right. \nonumber\\
&&\left.+I(\eta^{+},\eta^{+},\eta^{-})-I(\eta^{+},\eta^{-},\eta^{-})-I(\eta^{-},\eta^{-},\eta^{-})\right] \ . \label{diagram 2 integrals}
\end{eqnarray}

The contribution of the third diagram is given by
\begin{eqnarray}
{\mathcal G}_{3}^{(2)\delta^{2}}&=&16\delta^{2}g^{2}\!\!\int\!\frac{d^{4}pd^{4}kd^{4}\theta_{12}}{(2\pi)^8}\!\!\left[\!-\frac{1}{4}\bar{D}_{1}^{2}(p)\langle\Phi_{1}\bar{\Phi}_{0}\rangle\right]\!\!\left[\!-\frac{1}{4}D_{2}^{2}(k)\langle\bar{\Phi}_{1}\Phi_{1}\rangle\right]\!\!\left[\!\frac{1}{16}\bar{D}_{1}^{2}(q)D_{2}^{2}(-q)\langle\Phi_{0}\bar{\Phi}_{1}\rangle\right] \nonumber\\
&=&\frac{\delta^{2}g^{2}}{(16)^{3}}a^{2}b^{2}\int\!\frac{d^{4}pd^{4}k}{(2\pi)^8}C(p)C(q)\left\{E(k){\mathcal I}_{5}(\theta,\bar\theta)+\frac{1}{16}b^{2}B(k){\mathcal I}_{6}(\theta,\bar\theta)\right\} \ . \label{I3}
\end{eqnarray}
Plugging ${\mathcal I}_{5}(\theta,\bar\theta)$ and ${\mathcal I}_{6}(\theta,\bar\theta)$ into (\ref{I3}), we obtain:
\begin{equation}
{\mathcal G}_{3}^{(2)\delta^{2}}=16\delta^{2}g^{2}a^{2}b^{2}\!\left\{\!\int\!\frac{d^{4}pd^{4}k}{(2\pi)^8}C(p)E(k)C(q)p^{2}q^{2}\!+\!b^{2}\!\int\!\frac{d^{4}pd^{4}k}{(2\pi)^8}C(p)B(k)C(q)p^{2}q^{2}\!\right\} \ . \label{third diagram}
\end{equation}
Decomposing the momentum space integrals we have:
\begin{equation}
{\mathcal G}_{3}^{(2)\delta^{2}}=2\delta^{2}g^{2}a^{2}\left[I(\eta^{+},\eta^{+},\eta^{+})-I(\eta^{+},\eta^{+},\eta^{-})-I(\eta^{+},\eta^{-},\eta^{-})+I(\eta^{-},\eta^{-},\eta^{-})\right] \ . \label{diagram 3 integrals}
\end{equation}

The contribution of the fourth diagram is given by
\begin{eqnarray}
{\mathcal G}_{4}^{(2)\delta^{2}}&=&8\delta^{2}g^{2}\!\!\int\!\frac{d^{4}pd^{4}kd^{4}\theta_{12}}{(2\pi)^8}\!\!\left[\!-\frac{1}{4}\bar{D}_{1}^{2}(p)\langle\Phi_{1}\bar{\Phi}_{1}\rangle\right]\!\!\left[\!-\frac{1}{4}D_{2}^{2}(k)\langle\bar{\Phi}_{1}\Phi_{1}\rangle\right]\!\!\left[\!\frac{1}{16}\bar{D}_{1}^{2}(q)D_{2}^{2}(-q)\langle\Phi_{0}\bar{\Phi}_{0}\rangle\right] \nonumber\\
&=&\frac{8\delta^{2}g^{2}}{(16)^{2}}a^{2}b^{2}\int\!\frac{d^{4}pd^{4}k}{(2\pi)^8}B(q)\left\{E(p)E(k){\mathcal I}_{7}(\theta,\bar\theta)+\frac{1}{8}b^{2}E(p)B(k){\mathcal I}_{8}(\theta,\bar\theta)\right. \nonumber\\
&&\left.+\frac{1}{(16)^{2}}b^{4}B(p)B(k){\mathcal I}_{9}(\theta,\bar\theta)\right\} \nonumber\\
&&+\frac{8\delta^{2}g^{2}}{(16)^{2}}\int\!\frac{d^{4}pd^{4}k}{(2\pi)^8}A(q)(q^{2}+m^{2})\left\{E(p)E(k){\mathcal I}_{10}(\theta,\bar\theta)+\frac{1}{8}b^{2}E(p)B(k){\mathcal I}_{11}(\theta,\bar\theta)\right. \nonumber\\
&&\left.+\frac{1}{(16)^{2}}b^{4}B(p)B(k){\mathcal I}_{12}(\theta,\bar\theta)\right\} \ . \label{I4}
\end{eqnarray}
Plugging ${\mathcal I}_{7}(\theta,\bar\theta)$ - ${\mathcal I}_{12}(\theta,\bar\theta)$ into (\ref{I4}), we get:
\begin{eqnarray}
{\mathcal G}_{4}^{(2)\delta^{2}}&=&8\delta^{2}g^{2}b^{2}\left\{a^{2}\!\int\!\frac{d^{4}pd^{4}k}{(2\pi)^8}E(p)E(k)B(q)\!+\!2a^{2}b^{2}\!\int\!\frac{d^{4}pd^{4}k}{(2\pi)^8}E(p)B(k)B(q)\right. \nonumber\\
&&\left.+a^{2}b^{4}\!\int\!\frac{d^{4}pd^{4}k}{(2\pi)^8}B(p)B(k)B(q)\!-\!2\!\int\!\frac{d^{4}pd^{4}k}{(2\pi)^8}E(p)B(k)A(q)k^{2}q^{2}\right. \nonumber\\
&&\left.-b^{2}\!\int\!\frac{d^{4}pd^{4}k}{(2\pi)^8}B(p)B(k)A(q)q^{4}\!-\!2m^{2}\!\int\!\frac{d^{4}pd^{4}k}{(2\pi)^8}E(p)B(k)A(q)k^{2}\right. \nonumber\\
&&\left.-m^{2}b^{2}\!\int\!\frac{d^{4}pd^{4}k}{(2\pi)^8}B(p)B(k)A(q)q^{2}\right\} \ . \label{fourth diagram}
\end{eqnarray}
Decomposing the momentum space integrals, we obtain the total contribution of the fourth diagram:
\begin{eqnarray}
{\mathcal G}_{4}^{(2)\delta^{2}}&=&\delta^{2}g^{2}\left\{8a^{2}\left[-I(\eta^{2},\eta^{2},\eta^{2})+\frac{b}{\eta^{2}}I(\eta^{2},\eta^{2},\eta^{+})-\frac{b}{\eta^{2}}I(\eta^{2},\eta^{2},\eta^{-})\right]\right. \nonumber\\
&&\left.+8m^{2}\left[-2I(\eta^{2},\eta^{2},0)+\frac{\eta^+}{\eta^2}I(\eta^{2},\eta^{+},0)+\frac{\eta^-}{\eta^2}I(\eta^{2},\eta^{-},0)\right]\right. \nonumber\\
&&\left.+a^{2}\left[I(\eta^{+},\eta^{+},\eta^{+})+3I(\eta^{+},\eta^{+},\eta^{-})+3I(\eta^{+},\eta^{-},\eta^{-})+I(\eta^{-},\eta^{-},\eta^{-})\right]\right. \nonumber\\
&&\left.+2\left[-4J(\eta^{2},\eta^{2})+4J(\eta^{2},\eta^{+})+4J(\eta^{2},\eta^{-})\right.\right. \nonumber\\
&&\left.\left.-J(\eta^{+},\eta^{+})-2J(\eta^{+},\eta^{-})-J(\eta^{-},\eta^{-})\right]\right\} \ . \label{diagram 4 integrals}
\end{eqnarray}

Unlike the first three two-loop diagrams, ${\mathcal G}_{4}^{(2)\delta^{2}}$ gives a divergent contribution, 
however, its renormalization is trivial, since it is a vacuum diagram.


\subsection{Regularization and renormalization}

The divergent diagrams of order $\delta^2$ are:
\begin{eqnarray}
{\mathcal G}_{14}^{(1)\delta^2}&=&\left\{\frac{8\delta^{2}g^{2}}{\kappa}\left(\frac{1}{\epsilon}-\overline{\ln}\eta^{2}\right)\right. \nonumber\\
&&\left.+\frac{2\delta^{2}g^{2}}{\kappa b}\left[4b\overline{\ln}\eta^{2}\!-\!(\eta^{2}+2b)\overline{\ln}\eta^{+}\!+\!(\eta^{2}-2b)\overline{\ln}\eta^{-}\!+\!2b\right]\right\}\!\int\!d^{4}\theta\Phi_{0}\bar{\Phi}_{0} \ ; \label{G14 div}
\end{eqnarray}
\begin{eqnarray}
{\mathcal G}_{15}^{(1)\delta^2}&=&\left\{-\frac{2\delta^{2}gb}{\kappa}\left(\frac{1}{\epsilon}-\overline{\ln}\eta^{2}\right)\right. \nonumber\\
&&\left.-\frac{\delta^{2}gb}{2\kappa b}\!\left[4b\overline{\ln}\eta^{2}\!-\!(\eta^{2}\!+\!2b)\overline{\ln}\eta^{+}\!+\!(\eta^{2}\!-\!2b)\overline{\ln}\eta^{-}\!+\!2b\right]\right\}\!\!\int\!\!d^{4}\theta\bar{\theta}^{2}\Phi_{0}+h.c. \ ; \label{G15 div}
\end{eqnarray}
\begin{eqnarray}
{\mathcal G}_{16}^{(1)\delta^2}&=&\left\{\frac{\delta^{2}b^{2}}{2\kappa}\left(\frac{1}{\epsilon}-\overline{\ln}\eta^{2}\right)\right. \nonumber\\
&&\left.+\frac{\delta^{2}b^{2}}{8\kappa b}\left[4b\overline{\ln}\eta^{2}-(\eta^{2}+2b)\overline{\ln}\eta^{+}+(\eta^{2}-2b)\overline{\ln}\eta^{-}+2b\right]\right\}\int d^{4}\theta\theta^{2}\bar{\theta}^{2} \ . \label{G16 div}
\end{eqnarray}

The diagram ${\mathcal G}_{16}^{(1)\delta^2}$ is a vacuum diagram and thus its renormalization is trivial. To renormalize the divergent term in ${\mathcal G}_{14}^{(1)\delta^2}$ we introduce the counterterm
\begin{equation}
-\frac{8\delta^{2}g^{2}}{\kappa\epsilon}\int d^{4}\theta\Phi_{0R}\bar{\Phi}_{0R} \ , \label{CT2}
\end{equation}
and, for the divergent term in ${\mathcal G}_{15}^{(1)\delta^2}$, we introduce the counterterm
\begin{equation}
\frac{2\delta^{2}gb}{\kappa\epsilon}\int d^{4}\theta\bar{\theta}^{2}\Phi_{0R}+h.c.=\frac{2\delta^{2}gb}{\kappa\epsilon}\int d^{2}\theta\Phi_{0R}+h.c. \label{CT3}
\end{equation}

As mentioned in the previous section and showed in \cite{renormOR}, plugging the solution for the optimized 
parameter $b$ in (\ref{CT3}), only the K\"{a}hler potential is actually renormalized, in agreement with the 
nonrenormalization theorem.


\section{Summary of the Results and Numerical Analysis}

In this last section we summarize the perturbative results for the effective potential up to the order $\delta^2$. 
We also derive nonperturbative corrections to the effective potential by implementing the PMS criterion numerically. 
Up to this order,  the renormalized effective potential  can be written as 
\begin{equation}
{\mathcal V}_{eff}={\mathcal V}_{eff}^{tree}+{\mathcal V}_{eff}^{\delta^0}+{\mathcal V}_{eff}^{\delta^1}
+{\mathcal V}_{eff}^{\delta^2} \ . 
\label{vefffull}
\end{equation}

\noindent Below, we write separately the terms of the effective potential ${\mathcal V}_{eff}$ as follows:

The tree level potential is given by:
\begin{equation}
{\mathcal V}_{eff}^{tree}=V(\varphi_{1})=(\xi+g\varphi_{1}^{2})^{2}+m^{2}\varphi_{1}^{2} \ , \label{tree}
\end{equation}
with $\xi<0$.

The vacuum diagram of order $\delta^0$ is:
\begin{eqnarray}
{\mathcal V}_{eff}^{\delta^0}&=&\frac{1}{(4\pi)^{2}}\left\{\frac{1}{4}\!\left(M^{2}\!+\!a^{2}\right)^{2}\ln\!\left[1-\frac{b^{2}}{\left(M^{2}+a^{2}\right)^{2}}\right]\right. \nonumber\\
&&\left.+\frac{b}{2}\!\left(M^{2}\!+\!a^{2}\right)\!\ln\!\left[\frac{M^{2}+a^{2}+b}{M^{2}+a^{2}-b}\right]\!+\!\frac{b^{2}}{4}\ln\!\left[\frac{\left(M^{2}\!+\!a^{2}\right)^{2}-b^{2}}{\mu^4}\right]\!-\!\frac{3b^2}{4}\right\} \ . \label{bolha}
\end{eqnarray}

The one-loop contribution for the effective potential at $\mathcal{O}(\delta^1)$ is given by:
\begin{eqnarray}
{\mathcal V}_{eff}^{\delta^1}&=&\frac{\delta}{(4\pi)^{2}}\!\left\{b(b-4g\langle F_{0}\rangle)+2\left[a(a-4g\langle\varphi_{1}\rangle)+\rho M\right]\!\left(M^{2}+a^{2}\right)\overline{\ln}\left[M^{2}+a^{2}\right]\right. \nonumber\\
&&\!\left.+\left[a(4g\langle\varphi_{1}\rangle\!-\!a)\!+\!\frac{1}{2}(4g\langle F_{0}\rangle\!-\!b)\!-\!\rho M\right]\!\!\left(M^{2}\!+\!a^{2}\!+\!b\right)\overline{\ln}\left[M^{2}\!+\!a^{2}\!+\!b\right]\right. \nonumber\\
&&\!\left.+\left[a(4g\langle\varphi_{1}\rangle\!-\!a)\!-\!\frac{1}{2}(4g\langle F_{0}\rangle\!-\!b)\!-\!\rho M\right]\!\!\left(M^{2}\!+\!a^{2}\!-\!b\right)\overline{\ln}\left[M^{2}\!+\!a^{2}\!-\!b\right]\!\right\}. \label{1loop-delta1}
\end{eqnarray}

At the $\mathcal{O}(\delta^2)$, we separate the contributions of one- and two-loop diagrams:
\begin{equation}
{\mathcal V}_{eff}^{\delta^2}={\mathcal V}_{eff \, (I)}^{\delta^2}+{\mathcal V}_{eff \, (II)}^{\delta^2} \ ,
\end{equation}
where
\begin{eqnarray}
{\mathcal V}_{eff \, (I)}^{\delta^2}&=&\frac{\delta^2}{(4\pi)^{2}}\left\{\left[(a-4g\langle\varphi_{1}
\rangle)^{2}\left(M^{2}+3a^{2}\right)+a\rho M(a-4g\langle\varphi_{1}\rangle)+\rho^{2}\left(3M^{2}
+a^{2}\right)\right]\frac{}{}\right. \nonumber\\
&&\times\overline{\ln}\left(M^{2}+a^{2}\right) \nonumber\\
&&-\left.\frac{1}{2}\left[(a-4g\langle\varphi_{1}\rangle)\left[(a-4g\langle\varphi_{1}\rangle)(M^{2}+3a^{2}+b)
+2a(b-4g\langle F_{0}\rangle)+4a\rho M\right]\frac{}{}\right.\right. \nonumber\\
&&+\left.\left.\frac{1}{2}(b-4g\langle F_{0}\rangle)\left[(b-4g\langle F_{0}\rangle)+4\rho M\right]
+\rho^{2}(3M^{2}+a^{2}+b)\right]\overline{\ln}\left(M^{2}+a^{2}+b\right)\right. \nonumber\\
&&-\left.\frac{1}{2}\left[(a-4g\langle\varphi_{1}\rangle)\left[(a-4g\langle\varphi_{1}\rangle)(M^{2}+3a^{2}-b)
-2a(b-4g\langle F_{0}\rangle)+4a\rho M\right]\right.\right. \nonumber\\
&&+\left.\left.\frac{1}{2}(b-4g\langle F_{0}\rangle)\left[(b-4g\langle F_{0}\rangle)-4\rho M\right]
+\rho^{2}(3M^{2}+a^{2}-b)\right] \right.\nonumber\\
&&\times\left.\overline{\ln}\left(M^{2}+a^{2}-b\right)\right\} \ , \label{1loopdelta2}
\end{eqnarray}
and
\begin{eqnarray}
{\mathcal V}_{eff \, (II)}^{\delta^2}&=&\frac{\delta^{2}g^{2}}{(4\pi)^{4}}\left\{\frac{4}{(M^{2}+a^{2})}\left[4(M^{2}+a^{2})^{3}+b(M^{2}+a^{2}-b)(M^{2}-2a^{2})\right]\overline{\ln}^{2}(M^{2}+a^{2})\right. \nonumber\\
&&\left.-\frac{(M^{2}+a^{2}+b)}{(M^{2}+a^{2})}\left[(M^{2}+a^{2})(6M^{2}+17a^{2})+2b(M^{2}+3a^{2})\right]\overline{\ln}^{2}(M^{2}+a^{2}+b)\right. \nonumber\\
&&\left.-3(M^{2}+a^{2}-b)(2M^{2}+3a^{2}-2b)\overline{\ln}^{2}(M^{2}+a^{2}-b)\right. \nonumber\\
&&\left.+8(M^{2}+a^{2}+b)(2a^{2}-b)\overline{\ln}(M^{2}+a^{2})\overline{\ln}(M^{2}+a^{2}+b)\right. \nonumber\\
&&\left.+\frac{8a^{2}b(M^{2}+a^{2}-b)}{(M^{2}+a^{2})}\overline{\ln}(M^{2}+a^{2})\overline{\ln}(M^{2}+a^{2}-b)\right. \nonumber\\
&&\left.-2(M^{2}+a^{2}+b)(2M^{2}+3a^{2}-2b)\overline{\ln}(M^{2}+a^{2}+b)\overline{\ln}(M^{2}+a^{2}-b)\right. \nonumber\\
&&\left.+\frac{8M^{2}b}{(M^{2}\!+\!a^{2})}\!\left[b\ln\left(\frac{(M^{2}\!+\!a^{2})^{2}}{(M^{2}\!+\!a^{2})^{2}\!-\!b^{2}}\right)\!-\!(M^{2}\!+\!a^{2})\ln\left(\frac{M^{2}\!+\!a^{2}\!+\!b}{M^{2}\!+\!a^{2}\!-\!b}\right)\right]\!\overline{\ln}b\right. \nonumber\\
&&\left.-24(M^{2}+a^{2})(2M^{2}+3a^{2})\overline{\ln}(M^{2}+a^{2})\right. \nonumber\\
&&\left.+12(M^{2}+a^{2}+b)(2M^{2}+3a^{2}+2b)\overline{\ln}(M^{2}+a^{2}+b)\right. \nonumber\\
&&\left.+12(M^{2}+a^{2}-b)(2M^{2}+3a^{2}-2b)\overline{\ln}(M^{2}+a^{2}-b)\right. \nonumber\\
&&\left.+a^{2}\!\left[4\xi(M^{2}\!+\!a^{2},M^{2}\!+\!a^{2},M^{2}\!+\!a^{2})\!-\!3\xi(M^{2}\!+\!a^{2}\!+\!b,M^{2}\!+\!a^{2}\!+\!b,M^{2}\!+\!a^{2}\!+\!b)\right.\right. \nonumber\\
&&\left.\left. \ \ \ \ \ \ -\xi(M^{2}\!+\!a^{2}\!+\!b,M^{2}\!+\!a^{2}\!-\!b,M^{2}\!+\!a^{2}\!-\!b)\right]\right. \nonumber\\
&&\left.+\frac{4a^{2}(M^{2}+a^{2}-b)}{(M^{2}+a^{2})}\left[\xi(M^{2}\!
+\!a^{2},M^{2}\!+\!a^{2},M^{2}\!+\!a^{2}\!+\!b)\right.\right.\nonumber\\
&&-\left.\left.\xi(M^{2}\!+\!a^{2},M^{2}\!+\!a^{2},M^{2}\!+\!a^{2}\!-\!b)\right]\right. \nonumber\\
&&\left.+\frac{8M^{2}b}{(M^{2}+a^{2})}\left[(M^{2}+a^{2}+b)\mbox{Li}_{2}\!\left(\frac{M^{2}
+a^{2}}{M^{2}+a^{2}+b}\right)\right.\right.\nonumber\\
&&\left.\left.+(M^{2}+a^{2}-b)\mbox{Li}_{2}\!\left(\frac{M^{2}+a^{2}-b}{M^{2}+a^{2}}\right)\right]
-56b^{2}-\frac{4}{3}b\pi^{2}(2M^{2}+b)\right\} \ . \label{2loopdelta2}
\end{eqnarray}


\subsection{Optimization Procedure}

Let us now describe the numerical results obtained  by implementing the PMS criterion. Our numerical 
approach consists in solving a system of coupled equations, to determine the solutions for the optimal 
parameters $a$, $b$ and $\rho$ that satisfy the PMS criterion and minimize the effective potential 
up to the order $\delta^2$. We start comparing our numerical results with the analytical results derived at 
order $\delta^1$ and implement the optimization at order $\delta^2$.

The numerical implementation was performed using {\it Mathematica} \cite{math}. When we apply the 
PMS criterion to the effective potential there are three coupled equations to solve:
\begin{eqnarray}
\frac{\partial{\mathcal V}_{eff}}{\partial a}&=&0 \ , \nonumber \\
\frac{\partial{\mathcal V}_{eff}}{\partial b}&=&0 \ , \nonumber \\
\frac{\partial{\mathcal V}_{eff}}{\partial\rho}&=&0 \ .
\end{eqnarray}

We use the effective potential evaluated up to the order $\delta^2$, which is given by Eq. (\ref{vefffull}). 
The criterion that we have stablished to choose the optimum roots is to follow the same family of roots and 
work with the roots that minimize the effective potential. Although it is not possible to derive analytical 
solutions to the PMS equations up to the order $\delta^2$, it can be seen that the solutions can be written as
\begin{eqnarray}
a_{0}&=&4g\langle\varphi_{1}\rangle+\hbar A(g,\Phi,\bar{\Phi}) \ , \nonumber\\
b_{0}&=&4g\langle F_{0}\rangle+\hbar B(g,\Phi,\bar{\Phi}) \ , \nonumber \\
\rho_{0}&=&0+\hbar C(g,\Phi,\bar{\Phi}) \ , 
\end{eqnarray}
where $A$, $B$ and $C$ are corrections to the order $\delta^1$ solutions. This naturally brings nonlinear $g$ contributions 
and generates nonperturbative results that go beyond the one- and two-loop results derived in \cite{OFTdelta,background}. 

\begin{figure}[htb]
 \centerline{\psfig{file=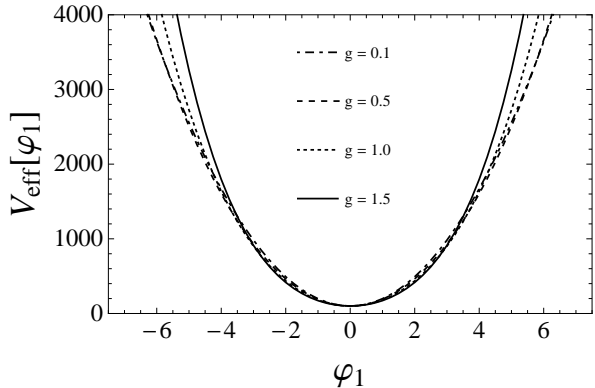,scale=0.665,angle=0}}
\caption{The results for tree level effective potential for different values of the coupling constant $g$. 
Parameters: $\xi = -10.0$, $m = 10.0$.}
\label{fig21}
\end{figure}

The parameters were chosen so that spontaneous supersymmetry breaking appears in the O'Raifeartaigh model. We 
choose $\xi = -10$ and $m = 10$ and perform one rescaling in all quantities in terms of the renormalization scale $\mu$ or, 
in other words, our quantities are given in unities of $\mu$. In Fig. \ref{fig21} we show the tree level effective potential for the 
O'Raifeartaigh model, and we can see that SUSY is spontaneously broken, since the value of the effective potential at the 
minimum is different from zero.

In a previous work~\cite{OFTdelta} was shown in detail that different optimization procedures FAC (Fastest Apparent Convergence)\footnote{for a description of 
the optimization procedure FAC please see Ref.~\cite{OFTdelta}} and PMS give the same result for the optimal parameters $a,\,\,b$ and $\rho$ 
at $\mathcal{O}(\delta^1)$,  and at this order was possible to implement these two optimization procedures analytically. 

\begin{figure}[ht]
\centerline{ 
\psfig{file=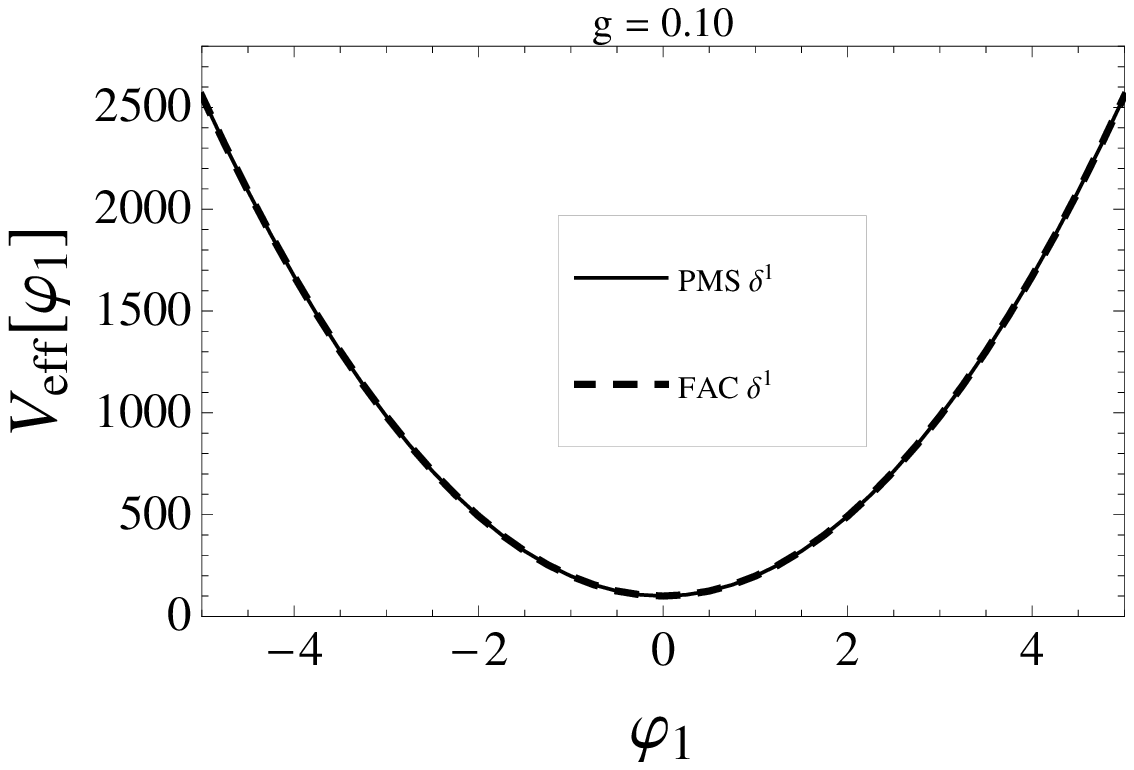,scale=0.565,angle=0}  
\psfig{file=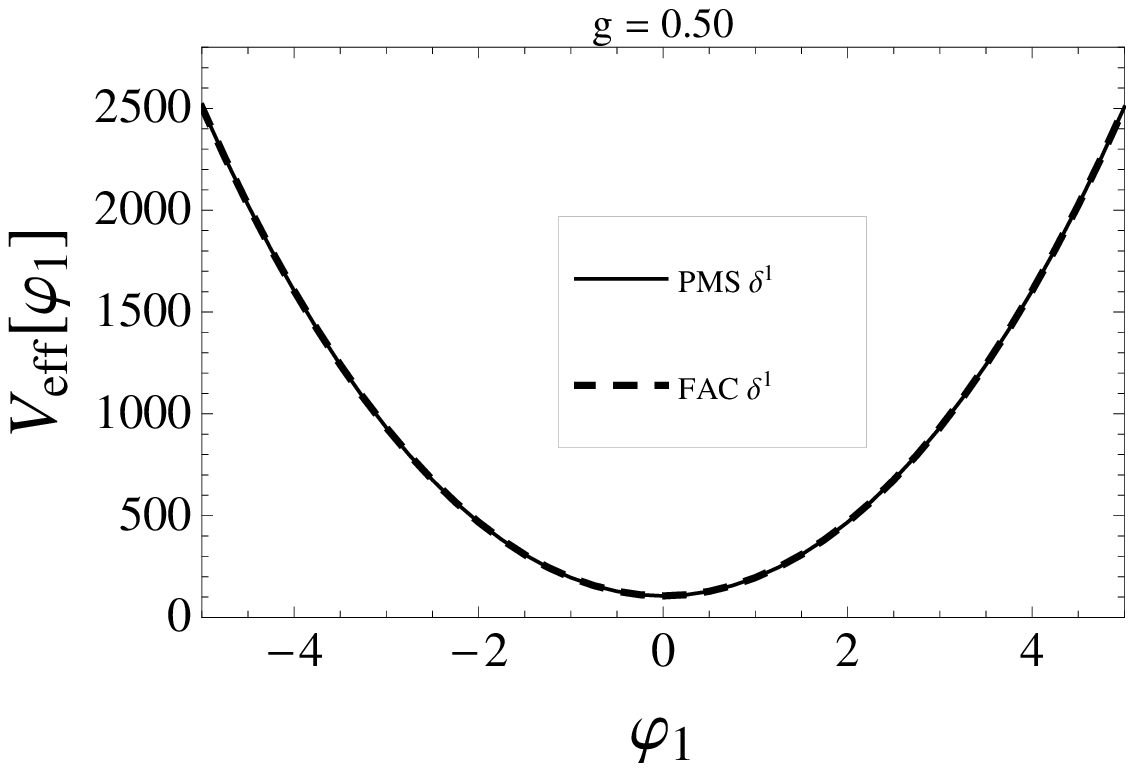,scale=0.565,angle=0}}
\centerline{ \psfig{file=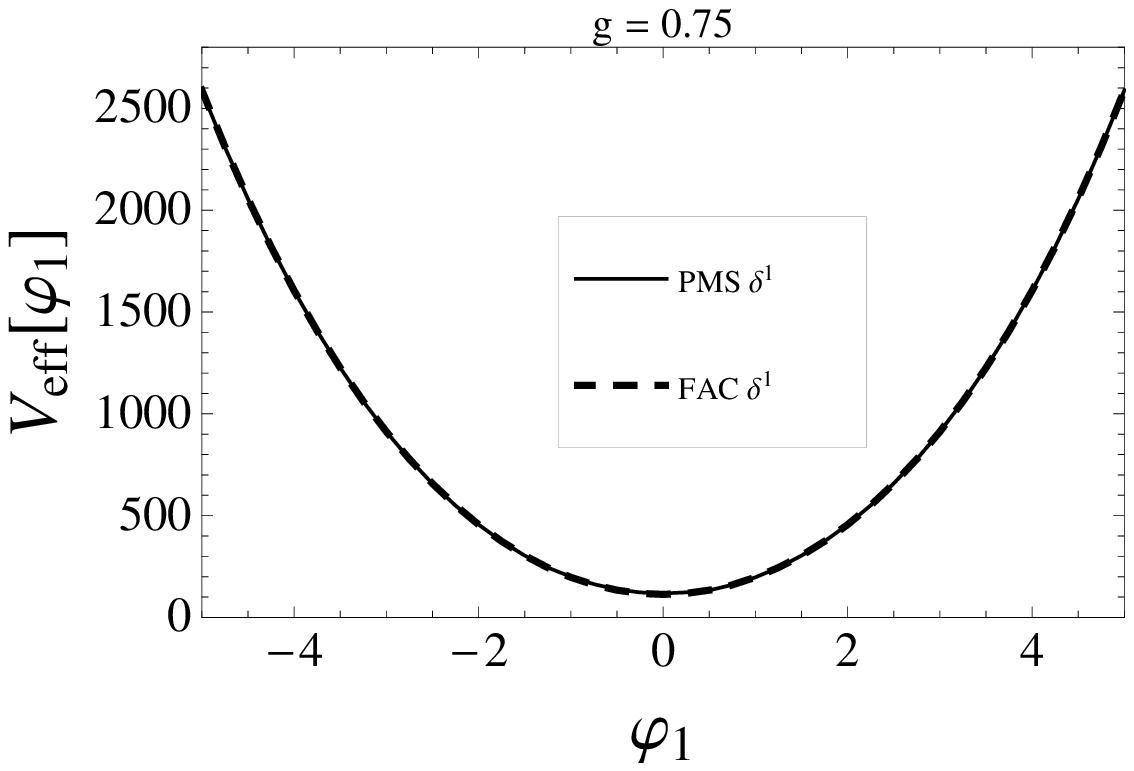,scale=0.565,angle=0}
\psfig{file=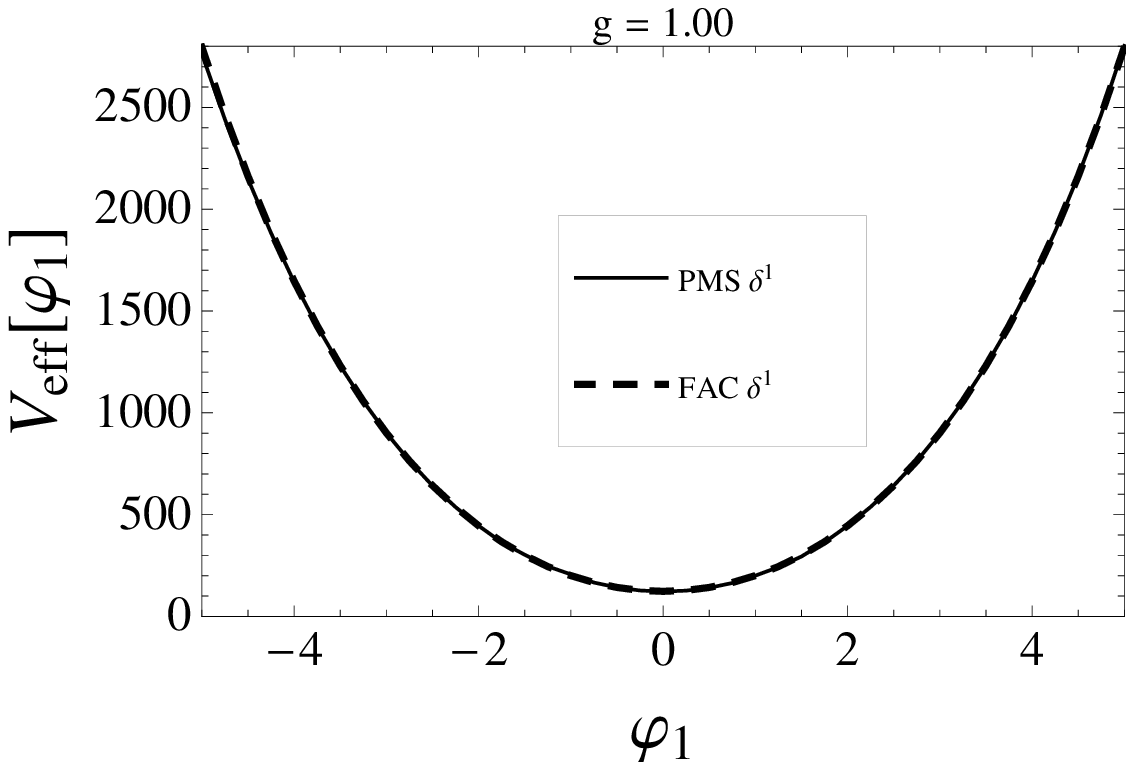,scale=0.565,angle=0}}
\caption{\sf The results for $V_{eff}(\varphi_1)$ using the optimization procedures FAC 
and PMS up to the $\mathcal{O}(\delta^1)$  in LDE for different values of the coupling constant 
$g$. Parameters: $\xi = -10.0$, $m = 10.0$.}
\label{fig22}
\end{figure}

The effective potential obtained in this case is shown in Fig. \ref{fig22}, where we show the 
results obtained up to the order $\delta^1$ using the PMS and FAC criteria. Our analytical and 
numerical results including only $\mathcal{O}(\delta^1)$ 
contributions to the effective potential have shown that the solution that minimizes the effective potential is $\rho=0$. In this 
figure we apply the PMS criterion numerically at $\mathcal{O}(\delta^1)$ and compare with the effective potential analytically 
evaluated with the FAC and PMS criteria. We change the value of the coupling constant $g$ and we can see that for different optimization 
procedures we obtain the same effective potential, as it should be, based on the analytical results obtained by the different 
optimization procedures.

We start with the comparison between the nonperturbative effective potential up to the $\mathcal{O}(\delta^1)$ 
and the $\mathcal{O}(\delta^2)$, allowing us to gauge the performance of each optimization procedure, regarding 
both reliability and indications of the convergence of the method. 

\begin{table}[!htb]
\begin{tabular}{c|c|c|c}
\hline \hline
& \multicolumn{3}{c}{$V_{\rm{eff}}\left[\varphi_1=0\right]$ }\\ \hline     
& \multicolumn{1}{c|}{PMS $\mathcal{O}(\delta^1)$ }   &  \multicolumn{1}{c|}
{FAC $\mathcal{O}(\delta^1)$}  & \multicolumn{1}{c}{PMS $\mathcal{O}(\delta^2)$}     \\ \hline
$g=0.01$  & 100.002      &100.002    &  100.002              \\ \hline
$g=0.10$  & 100.233     & 100.233    &  100.231             \\ \hline
$g=0.50$  & 105.828     & 105.828    &  104.544              \\ \hline
$g=0.75$  & 113.101     & 113.101    &   107.394             \\ \hline
$g=1.00$  & 123.260    & 123.260    &  107.865             \\ \hline
$g=1.20$ &  133.448   & 133.448      &  105.631           \\  \hline \hline
\end{tabular}
\caption{Dependence of the minimum of effective potential with the coupling constant $g$ including the 
corrections at $\mathcal{O}(\delta^2)$. In this table we show the analytical results using FAC and numerical 
results for PMS. Parameters are the same of Fig. \ref{fig22}} 
\label{tabd2}
\end{table}

As a result, we find that at the $\mathcal{O}(\delta^2)$ the solution that minimizes the effective 
potential is again $\rho=0$. We can see in Table \ref{tabd2} the results for the value of the minimum of the effective potential, and we note that the results at the $\mathcal{O}(\delta^2)$ are very similar to 
the results at the $\mathcal{O}(\delta^1)$ up to $g\sim0.5$, indicating that for the parameters 
used here, $g\sim0.5$ can be a regime of strong coupling. To our knowledge this is the first work in the literature that performs the evaluation beyond the one-loop approximation of the effective potential for the O'Raifeartaigh model (with spontaneous SUSY breaking). For the parameters considered in this study we can see strong indications that the nonperturbative method of LDE is appropriate to deal with the O'Raifeartaigh 
model. The results at the $\mathcal{O}(\delta^1)$ are the same as those at the $\mathcal{O}(\delta^2)$ up to some value of the coupling constant, indicating some convergence of the 
nonperturbative method.

\section{Concluding Remarks}

The investigation we have pursued in this paper and the results of our explicit supergraph evaluations confirm that superspace techniques, even if SUSY is spontaneously or explicitly broken, can be consistently combined with the LDE approach to compute higher order corrections to effective potentials in the framework of supersymmetric models. We just highlight that we are here bound to the minimal O'Raifeartaigh model, but the extension to generalized (Wess-Zumino type) matter models can naturally be pushed forward. Fayet-Iliopoulos D-term breaking models are also very interesting to be reassessed with the approach we have adopted in the present work. It is true that these first attempts to go through higher orders with the LDE procedure in the realm of supersymmetric models should drive us to explicit higher order calculations to compute corrections in the MSSM, which may allow us to use our semi-nonpeturbative results in connection with the constraints on SUSY as imposed by the phenomenology of the LHC/ATLAS and CMS collaborations.

The level of convergence of our results is satisfactory and, since our category of SUSY spontaneously broken model is still protected by the SUSY nonrenormalization theorem (the spontaneous breaking is a soft mechanism), we do not run into troubles with our perturbative calculations, for our coupling constants do not risk to take us to the strong coupling regime. This very same point must be reconsidered if we are dealing with a supersymmetric gauge theory, as it is the case of the MSSM, which is the ultimate framework to connect SUSY with the accessible energies. Before going directly to the MSSM, we intend to extend the calculations we have performed here to include the Fayet-Iliopoulos D-term~\cite{Fayet-Iliopoulos} models, in which SUSY is broken and gauge symmetry may also be. In this case, there appears a number of non-trivial aspects in connection with superspace and supergraph calculations, such as the gauge choice (unitary or 't Hooft's gauge choices in superspace) and a rich structure of $\theta$-dependent terms~\cite{Girardello-Grisaru} in the sector of gauge superfield propagators~\cite{Helayel 2}. The consideration of the Fayet-Iliopoulos models is clearly mandatory as a step prior to extend our analysis to the physics of the MSSM. We shall be next focusing on this specific step of our project.

Physics in three-dimensional space-time has been acquiring special interest, in view of a very rich diversity of lower-dimensional Condensed Matter (CM) systems that can be approached by quantum field-theoretic methods. On the top of that, more recently, renormalizable and unitary massive gravity models has driven the attention to planar gravity models. We know that SUSY may be connected to both types of systems, CM and gravity. Now, the realization of ${\mathcal N}=1$ SUSY breaking in three space-time dimensions is very special, since the structure underneath it is real and not complex (as it happens in ${\mathcal N}=1, \ D=4$ SUSY  or ${\mathcal N}=2, \ D=3$ SUSY). Also, renormalizability allows a higher-power scalar potential in $D=3$, so that F-term SUSY breaking demands reassessment and a number of peculiarities show up. Also, a Fayet-Iliopoulos term is not just like in ${\mathcal N}=1, \ D=4$ or ${\mathcal N}=2, \ D=3$ SUSY, for the gauge potential multiplet is spinorial in simple $3D$-SUSY. So, in view of these nontrivial aspects and the potentialities of supersymmetric planar systems, we believe that there should be some interest in re-analyzing the methods and re-discussing the results of the present work.

\acknowledgments

M. C. B. Abdalla and Daniel L. Nedel would like to thank CNPq, grants 306276/2009-7 and 501317/2009-0, for financial support. J. A. 
Helay\"{e}l-Neto would like to thank CNPq and FAPERJ for constant support. R.L.S.F. would like to thank FAPEMIG for financial support and R. O. 
Ramos and M. B. Pinto for discussions on related matters. Carlos R. Senise Jr. thanks CAPES-Brazil and Programa Rec\'em-Doutor-UNESP 
for financial support.

\section*{Appendix: Superspace Integrals}

In this appendix, the superspace integrals appearing in section III are listed.

The integrals appearing in the one-loop diagrams of ${\mathcal O}(\delta^2)$ are: 
\begin{eqnarray}
{\mathcal J}_{1}(\theta,\bar\theta)&=&\int d^{4}\theta_{12}\Phi_{0}(1)\Phi_{0}(2)\left[\bar{D}_{1}^{2}(k)\bar{\theta}_{1}^{2}D_{1}^{2}(k)\delta^{4}_{12}\right]\left[\bar{D}_{2}^{2}(-k)\bar{\theta}_{2}^{2}D_{2}^{2}(-k)\delta^{4}_{12}\right] \nonumber\\
&=&(16)^{2}\langle F_{0}\rangle^{2} \ ; \\
&& \nonumber\\
{\mathcal J}_{2}(\theta,\bar\theta)&=&\int d^{4}\theta_{12}\Phi_{0}(1)\left[\bar{D}_{1}^{2}(k)\bar{\theta}_{1}^{2}D_{1}^{2}(k)\delta^{4}_{12}\right]\left[\theta_{2}^{2}\bar{D}_{2}^{2}(-k)\bar{\theta}_{2}^{2}D_{2}^{2}(-k)\delta^{4}_{12}\right] \nonumber\\
&=&(16)^{2}\langle F_{0}\rangle \ ; \\
&& \nonumber\\
{\mathcal J}_{3}(\theta,\bar\theta)&=&\int d^{4}\theta_{12}\left[\theta_{1}^{2}\bar{D}_{1}^{2}(k)\bar{\theta}_{1}^{2}D_{1}^{2}(k)\delta^{4}_{12}\right]\left[\theta_{2}^{2}\bar{D}_{2}^{2}(-k)\bar{\theta}_{2}^{2}D_{2}^{2}(-k)\delta^{4}_{12}\right] \nonumber\\
&=&(16)^{2} \ ; \\
&& \nonumber\\
{\mathcal J}_{4}(\theta,\bar\theta)&=&\int d^{4}\theta_{12}\Phi_{0}(1)\Phi_{1}(2)\left[\bar{D}_{1}^{2}(k)\bar{\theta}_{1}^{2}D_{1}^{2}(k)\delta^{4}_{12}\right]\left[\bar{D}_{2}^{2}(-k)D_{2}^{2}(-k)\delta^{4}_{12}\right] \nonumber\\
&=&-(16)^{2}k^{2}\langle F_{0}\rangle\langle\varphi_{1}\rangle \ ; \\
&& \nonumber\\
{\mathcal J}_{5}(\theta,\bar\theta)&=&\int d^{4}\theta_{12}\Phi_{0}(1)\Phi_{1}(2)\left[\bar{D}_{1}^{2}(k)\bar{\theta}_{1}^{2}D_{1}^{2}(k)\delta^{4}_{12}\right]\left[\bar{D}_{2}^{2}(-k)\theta_{2}^{2}\bar{\theta}_{2}^{2}D_{2}^{2}(-k)\delta^{4}_{12}\right] \nonumber\\
&=&(16)^{2}\langle F_{0}\rangle\langle\varphi_{1}\rangle \ ; \\
&& \nonumber\\
{\mathcal J}_{6}(\theta,\bar\theta)&=&\int d^{4}\theta_{12}\Phi_{0}(1)\left[\bar{D}_{1}^{2}(k)\bar{\theta}_{1}^{2}D_{1}^{2}(k)\delta^{4}_{12}\right]\left[\bar{D}_{2}^{2}(-k)D_{2}^{2}(-k)\delta^{4}_{12}\right] \nonumber\\
&=&-(16)^{2}k^{2}\langle F_{0}\rangle \ ; \\
&& \nonumber\\
{\mathcal J}_{7}(\theta,\bar\theta)&=&\int d^{4}\theta_{12}\Phi_{1}(1)\left[\bar{D}_{1}^{2}(k)\bar{\theta}_{1}^{2}D_{1}^{2}(k)\delta^{4}_{12}\right]\left[\theta_{2}^{2}\bar{D}_{2}^{2}(-k)D_{2}^{2}(-k)\delta^{4}_{12}\right] \nonumber\\
&=&-(16)^{2}k^{2}\langle\varphi_{1}\rangle \ ; \\
&& \nonumber\\
{\mathcal J}_{8}(\theta,\bar\theta)&=&\int d^{4}\theta_{12}\Phi_{1}(1)\left[\bar{D}_{1}^{2}(k)\bar{\theta}_{1}^{2}D_{1}^{2}(k)\delta^{4}_{12}\right]\left[\theta_{2}^{2}\bar{D}_{2}^{2}(-k)D_{2}^{2}(-k)\theta_{2}^{2}\bar{\theta}_{2}^{2}\delta^{4}_{12}\right] \nonumber\\
&=&(16)^{2}\langle\varphi_{1}\rangle \ ; \\
&& \nonumber\\
{\mathcal J}_{9}(\theta,\bar\theta)&=&\int d^{4}\theta_{12}\left[\theta_{1}^{2}\bar{D}_{1}^{2}(k)\bar{\theta}_{1}^{2}D_{1}^{2}(k)\delta^{4}_{12}\right]\left[\bar{D}_{2}^{2}(-k)D_{2}^{2}(-k)\delta^{4}_{12}\right] \nonumber\\
&=&-(16)^{2}k^{2} \ ; \\
&& \nonumber\\
{\mathcal J}_{10}(\theta,\bar\theta)&=&\int d^{4}\theta_{12}\Phi_{1}(1)\Phi_{1}(2)\left[\bar{D}_{1}^{2}(k)\bar{\theta}_{1}^{2}D_{1}^{2}(k)\delta^{4}_{12}\right]\left[\bar{D}_{2}^{2}(-k)D_{2}^{2}(-k)\theta_{2}^{2}\delta^{4}_{12}\right] \nonumber\\
&=&-(16)^{2}k^{2}\langle\varphi_{1}\rangle^{2} \ ; \\
&& \nonumber\\
{\mathcal J}_{11}(\theta,\bar\theta)&=&\int d^{4}\theta_{12}\Phi_{1}(1)\Phi_{1}(2)\left[\bar{D}_{1}^{2}(k)D_{1}^{2}(k)\delta^{4}_{12}\right]\left[\bar{D}_{2}^{2}(-k)D_{2}^{2}(-k)\delta^{4}_{12}\right] \nonumber\\
&=&0 \ ; \\
&& \nonumber\\
{\mathcal J}_{12}(\theta,\bar\theta)&=&\int d^{4}\theta_{12}\Phi_{1}(1)\Phi_{1}(2)\left[\bar{D}_{1}^{2}(k)D_{1}^{2}(k)\theta_{1}^{2}\bar{\theta}_{1}^{2}\delta^{4}_{12}\right]\left[\bar{D}_{2}^{2}(-k)D_{2}^{2}(-k)\theta_{2}^{2}\bar{\theta}_{2}^{2}\delta^{4}_{12}\right] \nonumber\\
&=&(16)^{2}\langle\varphi_{1}\rangle^{2} \ ; \\
&& \nonumber\\
{\mathcal J}_{13}(\theta,\bar\theta)&=&\int d^{4}\theta_{12}\Phi_{1}(1)\left[\bar{D}_{1}^{2}(k)D_{1}^{2}(k)\delta^{4}_{12}\right]\left[\bar{D}_{2}^{2}(-k)D_{2}^{2}(-k)\delta^{4}_{12}\right] \nonumber\\
&=&0 \ ; \\
&& \nonumber\\
{\mathcal J}_{14}(\theta,\bar\theta)&=&\int d^{4}\theta_{12}\left[\bar{D}_{1}^{2}(k)D_{1}^{2}(k)\delta^{4}_{12}\right]\left[\bar{D}_{2}^{2}(-k)D_{2}^{2}(-k)\delta^{4}_{12}\right] \nonumber\\
&=&0 \ ; \\
&& \nonumber\\
{\mathcal J}_{15}(\theta,\bar\theta)&=&\int d^{4}\theta_{12}\Phi_{0}(1)\bar{\Phi}_{0}(2)\left[\bar{D}_{1}^{2}(k)\delta^{4}_{12}\right]\left[D_{2}^{2}(-k)\delta^{4}_{12}\right] \nonumber\\
&=&16\langle F_{0}\rangle^{2} \ ; \\
&& \nonumber\\
{\mathcal J}_{16}(\theta,\bar\theta)&=&\int d^{4}\theta_{12}\Phi_{0}(1)\bar{\Phi}_{0}(2)\left[\bar{D}_{1}^{2}(k)\delta^{4}_{12}\right]\left[D_{2}^{2}(-k)\bar{D}_{2}^{2}(-k)\bar{\theta}_{2}^{2}\theta_{2}^{2}D_{2}^{2}(-k)\delta^{4}_{12}\right] \nonumber\\
&=&(16)^{2}\langle F_{0}\rangle^{2} \ ; \\
&& \nonumber\\
{\mathcal J}_{17}(\theta,\bar\theta)&=&\int d^{4}\theta_{12}\Phi_{0}(1)\bar{\Phi}_{0}(2)\left[\bar{D}_{1}^{2}(k)D_{1}^{2}(k)\theta_{1}^{2}\bar{\theta}_{1}^{2}\bar{D}_{1}^{2}(k)\delta^{4}_{12}\right]\left[D_{2}^{2}(-k)\bar{D}_{2}^{2}(-k)\bar{\theta}_{2}^{2}\theta_{2}^{2}D_{2}^{2}(-k)\delta^{4}_{12}\right] \nonumber\\
&=&(16)^{3}\langle F_{0}\rangle^{2} \ ; \\
&& \nonumber\\
{\mathcal J}_{18}(\theta,\bar\theta)&=&\int d^{4}\theta_{12}\Phi_{0}(1)\left[\bar{D}_{1}^{2}(k)\delta^{4}_{12}\right]\left[\bar{\theta}_{2}^{2}D_{2}^{2}(-k)
\delta^{4}_{12}\right] \nonumber\\
&=&16\langle F_{0}\rangle \ ; \\
&& \nonumber\\
{\mathcal J}_{19}(\theta,\bar\theta)&=&\int d^{4}\theta_{12}\Phi_{0}(1)\left[\bar{D}_{1}^{2}(k)D_{1}^{2}(k)\theta_{1}^{2}\bar{\theta}_{1}^{2}
\bar{D}_{1}^{2}(k)\delta^{4}_{12}\right]\left[\bar{\theta}_{2}^{2}D_{2}^{2}(-k)\bar{D}_{2}^{2}(-k)\bar{\theta}_{2}^{2}\theta_{2}^{2}D_{2}^{2}(-k)
\delta^{4}_{12}\right] \nonumber\\
&=&(16)^{3}\langle F_{0}\rangle \ ; \\
&& \nonumber\\
{\mathcal J}_{20}(\theta,\bar\theta)&=&\int d^{4}\theta_{12}\left[\theta_{1}^{2}\bar{D}_{1}^{2}(k)\delta^{4}_{12}\right]\left[\bar{\theta}_{2}^{2}D_{2}^{2}(-k)\delta^{4}_{12}\right] \nonumber\\
&=&16 \ ; \\
&& \nonumber\\
{\mathcal J}_{21}(\theta,\bar\theta)&=&\int d^{4}\theta_{12}\left[\theta_{1}^{2}\bar{D}_{1}^{2}(k)D_{1}^{2}(k)\theta_{1}^{2}\bar{\theta}_{1}^{2}\bar{D}_{1}^{2}(k)\delta^{4}_{12}\right]\left[\bar{\theta}_{2}^{2}D_{2}^{2}(-k)\bar{D}_{2}^{2}(-k)\bar{\theta}_{2}^{2}\theta_{2}^{2}D_{2}^{2}(-k)\delta^{4}_{12}\right] \nonumber\\
&=&(16)^{3} \ ; \\
&& \nonumber\\
{\mathcal J}_{22}(\theta,\bar\theta)&=&\int d^{4}\theta_{12}\Phi_{0}(1)\bar{\Phi}_{1}(2)\left[\bar{D}_{1}^{2}(k)\delta^{4}_{12}\right]\left[D_{2}^{2}(-k)\bar{D}_{2}^{2}(-k)D_{2}^{2}(-k)\delta^{4}_{12}\right] \nonumber\\
&=&-(16)^{2}k^{2}\langle F_{0}\rangle\langle\varphi_{1}\rangle \ ; \\
&& \nonumber\\
{\mathcal J}_{23}(\theta,\bar\theta)&=&\int d^{4}\theta_{12}\Phi_{0}(1)\bar{\Phi}_{1}(2)\left[\bar{D}_{1}^{2}(k)D_{1}^{2}(k)\theta_{1}^{2}\bar{\theta}_{1}^{2}\bar{D}_{1}^{2}(k)\delta^{4}_{12}\right]\left[D_{2}^{2}(-k)\bar{D}_{2}^{2}(-k)D_{2}^{2}(-k)\bar{\theta}_{2}^{2}\delta^{4}_{12}\right] \nonumber\\
&=&-(16)^{3}k^{2}\langle F_{0}\rangle\langle\varphi_{1}\rangle \ ; \\
&& \nonumber\\
{\mathcal J}_{24}(\theta,\bar\theta)&=&\int d^{4}\theta_{12}\Phi_{0}(1)\left[\bar{D}_{1}^{2}(k)D_{1}^{2}(k)\theta_{1}^{2}\bar{\theta}_{1}^{2}\bar{D}_{1}^{2}(k)\delta^{4}_{12}\right]\left[D_{2}^{2}(-k)\bar{D}_{2}^{2}(-k)D_{2}^{2}(-k)\bar{\theta}_{2}^{2}\delta^{4}_{12}\right] \nonumber\\
&=&-(16)^{3}k^{2}\langle F_{0}\rangle \ ; \\
&& \nonumber\\
{\mathcal J}_{25}(\theta,\bar\theta)&=&\int d^{4}\theta_{12}\bar{\Phi}_{1}(1)\left[D_{1}^{2}(k)\delta^{4}_{12}\right]\left[\theta_{2}^{2}\bar{D}_{2}^{2}(-k)\bar{\theta}_{2}^{2}D_{2}^{2}(-k)\bar{D}_{2}^{2}(-k)\delta^{4}_{12}\right] \nonumber\\
&=&-(16)^{2}k^{2}\langle\varphi_{1}\rangle \ ; \\
&& \nonumber\\
{\mathcal J}_{26}(\theta,\bar\theta)&=&\int d^{4}\theta_{12}\bar{\Phi}_{1}(1)\left[D_{1}^{2}(k)\bar{D}_{1}^{2}(k)\bar{\theta}_{1}^{2}\theta_{1}^{2}D_{1}^{2}(k)\delta^{4}_{12}\right]\left[\theta_{2}^{2}\bar{D}_{2}^{2}(-k)\bar{\theta}_{2}^{2}D_{2}^{2}(-k)\bar{D}_{2}^{2}(-k)\delta^{4}_{12}\right] \nonumber\\
&=&-(16)^{3}k^{2}\langle\varphi_{1}\rangle \ ; \\
&& \nonumber\\
{\mathcal J}_{27}(\theta,\bar\theta)&=&\int d^{4}\theta_{12}\left[\theta_{1}^{2}\bar{D}_{1}^{2}(k)D_{1}^{2}(k)\theta_{1}^{2}\bar{\theta}_{1}^{2}\bar{D}_{1}^{2}(k)\delta^{4}_{12}\right]\left[D_{2}^{2}(-k)\bar{D}_{2}^{2}(-k)D_{2}^{2}(-k)\bar{\theta}_{2}^{2}\delta^{4}_{12}\right] \nonumber\\
&=&-(16)^{3}k^{2} \ ; \\
&& \nonumber\\
{\mathcal J}_{28}(\theta,\bar\theta)&=&\int d^{4}\theta_{12}\Phi_{1}(1)\bar{\Phi}_{1}(2)\left[\bar{D}_{1}^{2}(k)\delta^{4}_{12}\right]\left[D_{2}^{2}(-k)\delta^{4}_{12}\right] \nonumber\\
&=&0 \ ; \\
&& \nonumber\\
{\mathcal J}_{29}(\theta,\bar\theta)&=&\int d^{4}\theta_{12}\Phi_{1}(1)\bar{\Phi}_{1}(2)\left[\bar{D}_{1}^{2}(k)\delta^{4}_{12}\right]\left[D_{2}^{2}(-k)\bar{\theta}_{2}^{2}\theta_{2}^{2}\delta^{4}_{12}\right] \nonumber\\
&=&16\langle\varphi_{1}\rangle^{2} \ ; \\
&& \nonumber\\
{\mathcal J}_{30}(\theta,\bar\theta)&=&\int d^{4}\theta_{12}\Phi_{1}(1)\bar{\Phi}_{1}(2)\left[\bar{D}_{1}^{2}(k)D_{1}^{2}(k)\theta_{1}^{2}
\bar{\theta}_{1}^{2}\bar{D}_{1}^{2}(k)\delta^{4}_{12}\right]\left[D_{2}^{2}(-k)\delta^{4}_{12}\right] \nonumber\\
&=&-(16)^{2}k^{2}\langle\varphi_{1}\rangle^{2} \ ; \\
&& \nonumber\\
{\mathcal J}_{31}(\theta,\bar\theta)&=&\int d^{4}\theta_{12}\Phi_{1}(1)\bar{\Phi}_{1}(2)\left[\bar{D}_{1}^{2}(k)D_{1}^{2}(k)\theta_{1}^{2}\bar{\theta}_{1}^{2}\bar{D}_{1}^{2}(k)\delta^{4}_{12}\right]\left[D_{2}^{2}(-k)\bar{\theta}_{2}^{2}\theta_{2}^{2}\delta^{4}_{12}\right] \nonumber\\
&=&(16)^{2}\langle\varphi_{1}\rangle^{2} \ ; \\
&& \nonumber\\
{\mathcal J}_{32}(\theta,\bar\theta)&=&\int d^{4}\theta_{12}\Phi_{1}(1)\left[\bar{D}_{1}^{2}(k)\delta^{4}_{12}\right]\left[D_{2}^{2}(-k)\bar{\theta}_{2}^{2}\theta_{2}^{2}\delta^{4}_{12}\right] \nonumber\\
&=&16\langle\varphi_{1}\rangle \ ; \\
&& \nonumber\\
{\mathcal J}_{33}(\theta,\bar\theta)&=&\int d^{4}\theta_{12}\left[\bar{D}_{1}^{2}(k)\delta^{4}_{12}\right]\left[D_{2}^{2}(-k)\delta^{4}_{12}\right] \nonumber\\
&=&0 \ ; \\
&& \nonumber\\
{\mathcal J}_{34}(\theta,\bar\theta)&=&\int d^{4}\theta_{12}\left[\bar{D}_{1}^{2}(k)\delta^{4}_{12}\right]\left[D_{2}^{2}(-k)\bar{\theta}_{2}^{2}\theta_{2}^{2}\delta^{4}_{12}\right] \nonumber\\
&=&16 \ ; \\
&& \nonumber\\
{\mathcal J}_{35}(\theta,\bar\theta)&=&\int d^{4}\theta_{12}\left[\bar{D}_{1}^{2}(k)D_{1}^{2}(k)\theta_{1}^{2}\bar{\theta}_{1}^{2}\bar{D}_{1}^{2}(k)\delta^{4}_{12}\right]\left[D_{2}^{2}(-k)\delta^{4}_{12}\right] \nonumber\\
&=&-(16)^{2}k^{2} \ ; \\
&& \nonumber\\
{\mathcal J}_{36}(\theta,\bar\theta)&=&\int d^{4}\theta_{12}\left[\bar{D}_{1}^{2}(k)D_{1}^{2}(k)\theta_{1}^{2}\bar{\theta}_{1}^{2}\bar{D}_{1}^{2}(k)\delta^{4}_{12}\right]\left[D_{2}^{2}(-k)\bar{\theta}_{2}^{2}\theta_{2}^{2}\delta^{4}_{12}\right] \nonumber\\
&=&(16)^{2} \ ; \\
&& \nonumber\\
{\mathcal J}_{37}(\theta,\bar\theta)&=&\int d^{4}\theta_{12}\Phi_{1}(1)\bar{\Phi}_{1}(2)\left[\bar{D}_{1}^{2}(k)\bar{\theta}_{1}^{2}D_{1}^{2}(k)\bar{D}_{1}^{2}(k)\delta^{4}_{12}\right]\left[D_{2}^{2}(-k)\theta_{2}^{2}\bar{D}_{2}^{2}(-k)D_{2}^{2}(-k)\delta^{4}_{12}\right] \nonumber\\
&=&(16)^{3}k^{4}\langle\varphi_{1}\rangle^{2} \ ; \\
&& \nonumber\\
{\mathcal J}_{38}(\theta,\bar\theta)&=&\int d^{4}\theta_{12}\Phi_{1}(1)\left[\bar{D}_{1}^{2}(k)\bar{\theta}_{1}^{2}D_{1}^{2}(k)\bar{D}_{1}^{2}(k)\delta^{4}_{12}\right]\left[D_{2}^{2}(-k)\theta_{2}^{2}\bar{D}_{2}^{2}(-k)D_{2}^{2}(-k)\delta^{4}_{12}\right] \nonumber\\
&=&(16)^{3}k^{4}\langle\varphi_{1}\rangle \ ; \\
&& \nonumber\\
{\mathcal J}_{39}(\theta,\bar\theta)&=&\int d^{4}\theta_{12}\left[\bar{D}_{1}^{2}(k)\bar{\theta}_{1}^{2}D_{1}^{2}(k)\bar{D}_{1}^{2}(k)\delta^{4}_{12}\right]\left[D_{2}^{2}(-k)\theta_{2}^{2}\bar{D}_{2}^{2}(-k)D_{2}^{2}(-k)\delta^{4}_{12}\right] \nonumber\\
&=&(16)^{3}k^{4} \ ; \\
&& \nonumber\\
{\mathcal J}_{40}(\theta,\bar\theta)&=&\int d^{4}\theta_{12}\Phi_{1}(1)\left[\bar{D}_{1}^{2}(k)\delta^{4}_{12}\right]\left[D_{2}^{2}(-k)\delta^{4}_{12}\right] \nonumber\\
&=&0 \ .
\end{eqnarray}

The integrals appearing in the two-loop diagrams of ${\mathcal O}(\delta^2)$ are:
\begin{eqnarray}
{\mathcal I}_{1}(\theta,\bar\theta)&=&\int\!d^{4}\theta_{12}\left[\bar{D}_{1}^{2}(p)D_{1}^{2}(p)\theta_{1}^{2}\delta_{12}^{4}\right]\left[\bar{D}_{2}^{2}(k)\bar{\theta}_{2}^{2}D_{2}^{2}(k)\delta_{12}^{4}\right]\left[\bar{D}_{1}^{2}(q)\bar{D}_{2}^{2}(-q)\bar{\theta}_{1}^{2}D_{1}^{2}(q)\delta_{12}^{4}\right] \nonumber\\
&=&4(16)^{3}p^{2} \ ; \\
&& \nonumber\\
{\mathcal I}_{2}(\theta,\bar\theta)&=&\int\!d^{4}\theta_{12}\left[\bar{D}_{1}^{2}(p)D_{1}^{2}(p)\delta_{12}^{4}\right]\left[\bar{D}_{2}^{2}(k)D_{2}^{2}(k)\delta_{12}^{4}\right]\left[\bar{D}_{1}^{2}(q)\bar{D}_{2}^{2}(-q)\bar{\theta}_{1}^{2}D_{1}^{2}(q)\delta_{12}^{4}\right] \nonumber\\
&=&0 \ ; \\
&& \nonumber\\
{\mathcal I}_{3}(\theta,\bar\theta)&=&\int\!d^{4}\theta_{12}\left[\bar{D}_{1}^{2}(p)D_{1}^{2}(p)\delta_{12}^{4}\right]\left[\bar{D}_{2}^{2}(k)\theta_{2}^{2}\bar{\theta}_{2}^{2}D_{2}^{2}(k)\delta_{12}^{4}\right]\left[\bar{D}_{1}^{2}(q)\bar{D}_{2}^{2}(-q)\bar{\theta}_{1}^{2}D_{1}^{2}(q)\delta_{12}^{4}\right] \nonumber\\
&=&4(16)^{3}p^{2} \ ; \\
&& \nonumber\\
{\mathcal I}_{4}(\theta,\bar\theta)&=&\int\!d^{4}\theta_{12}\left[\bar{D}_{1}^{2}(p)\theta_{1}^{2}\bar{\theta}_{1}^{2}D_{1}^{2}(p)\delta_{12}^{4}\right]\left[\bar{D}_{2}^{2}(k)\theta_{2}^{2}\bar{\theta}_{2}^{2}D_{2}^{2}(k)\delta_{12}^{4}\right]\left[\bar{D}_{1}^{2}(q)\bar{D}_{2}^{2}(-q)\bar{\theta}_{1}^{2}D_{1}^{2}(q)\delta_{12}^{4}\right] \nonumber\\
&=&-4(16)^{3} \ ; \\
&& \nonumber\\
{\mathcal I}_{5}(\theta,\bar\theta)&=&\int\!d^{4}\theta_{12}\left[\bar{D}_{1}^{2}(p)\bar{\theta}_{1}^{2}D_{1}^{2}(p)\bar{D}_{1}^{2}(p)\delta_{12}^{4}\right]\left[D_{2}^{2}(k)\delta_{12}^{4}\right]\left[\bar{D}_{1}^{2}(q)D_{2}^{2}(-q)D_{1}^{2}(q)\bar{D}_{1}^{2}(q)\theta_{1}^{2}\delta_{12}^{4}\right] \nonumber\\
&=&(16)^{4}p^{2}q^{2} \ ; \\
&& \nonumber\\
{\mathcal I}_{6}(\theta,\bar\theta)&=&\int\!d^{4}\theta_{12}\left[\bar{D}_{1}^{2}(p)\bar{\theta}_{1}^{2}D_{1}^{2}(p)\bar{D}_{1}^{2}(p)\delta_{12}^{4}\right]\left[D_{2}^{2}(k)\bar{D}_{2}^{2}(k)\bar{\theta}_{2}^{2}\theta_{2}^{2}D_{2}^{2}(k)\delta_{12}^{4}\right] \nonumber\\
&&\times\left[\bar{D}_{1}^{2}(q)D_{2}^{2}(-q)D_{1}^{2}(q)\bar{D}_{1}^{2}(q)\theta_{1}^{2}\delta_{12}^{4}\right] \nonumber\\
&=&(16)^{5}p^{2}q^{2} \ ; \\
&& \nonumber\\
{\mathcal I}_{7}(\theta,\bar\theta)&=&\int\!d^{4}\theta_{12}\left[\bar{D}_{1}^{2}(p)\delta_{12}^{4}\right]\left[D_{2}^{2}(k)\delta_{12}^{4}\right]\left[\bar{D}_{1}^{2}(q)D_{2}^{2}(-q)\theta_{1}^{2}\bar{\theta}_{1}^{2}\delta_{12}^{4}\right] \nonumber\\
&=&(16)^{2} \ ; \\
&& \nonumber\\
{\mathcal I}_{8}(\theta,\bar\theta)&=&\int\!d^{4}\theta_{12}\left[\bar{D}_{1}^{2}(p)\delta_{12}^{4}\right]\left[D_{2}^{2}(k)\bar{D}_{2}^{2}(k)\bar{\theta}_{2}^{2}\theta_{2}^{2}D_{2}^{2}(k)\delta_{12}^{4}\right]\left[\bar{D}_{1}^{2}(q)D_{2}^{2}(-q)\theta_{1}^{2}\bar{\theta}_{1}^{2}\delta_{12}^{4}\right] \nonumber\\
&=&(16)^{3} \ ; \\
&& \nonumber\\
{\mathcal I}_{9}(\theta,\bar\theta)&=&\int\!d^{4}\theta_{12}\left[\bar{D}_{1}^{2}(p)D_{1}^{2}(p)\theta_{1}^{2}\bar{\theta}_{1}^{2}\bar{D}_{1}^{2}(p)\delta_{12}^{4}\right]\left[D_{2}^{2}(k)\bar{D}_{2}^{2}(k)\bar{\theta}_{2}^{2}\theta_{2}^{2}D_{2}^{2}(k)\delta_{12}^{4}\right] \nonumber\\
&&\times\left[\bar{D}_{1}^{2}(q)D_{2}^{2}(-q)\theta_{1}^{2}\bar{\theta}_{1}^{2}\delta_{12}^{4}\right] \nonumber\\
&=&(16)^{4} \ ; \\
&& \nonumber\\
{\mathcal I}_{10}(\theta,\bar\theta)&=&\int\!d^{4}\theta_{12}\left[\bar{D}_{1}^{2}(p)\delta_{12}^{4}\right]\left[D_{2}^{2}(k)\delta_{12}^{4}\right]\left[\bar{D}_{1}^{2}(q)D_{2}^{2}(-q)\delta_{12}^{4}\right] \nonumber\\
&=&0 \ ; \\
&& \nonumber\\
{\mathcal I}_{11}(\theta,\bar\theta)&=&\int\!d^{4}\theta_{12}\left[\bar{D}_{1}^{2}(p)\delta_{12}^{4}\right]\left[D_{2}^{2}(k)\bar{D}_{2}^{2}(k)\bar{\theta}_{2}^{2}\theta_{2}^{2}D_{2}^{2}(k)\delta_{12}^{4}\right]\left[\bar{D}_{1}^{2}(q)D_{2}^{2}(-q)\delta_{12}^{4}\right] \nonumber\\
&=&-(16)^{3}k^{2} \ ; \\
&& \nonumber\\
{\mathcal I}_{12}(\theta,\bar\theta)&=&\int\!d^{4}\theta_{12}\left[\bar{D}_{1}^{2}(p)D_{1}^{2}(p)\theta_{1}^{2}\bar{\theta}_{1}^{2}\bar{D}_{1}^{2}(p)\delta_{12}^{4}\right]\left[D_{2}^{2}(k)\bar{D}_{2}^{2}(k)\bar{\theta}_{2}^{2}\theta_{2}^{2}D_{2}^{2}(k)\delta_{12}^{4}\right] \nonumber\\
&&\times\left[\bar{D}_{1}^{2}(q)D_{2}^{2}(-q)\delta_{12}^{4}\right] \nonumber\\
&=&-(16)^{4}q^{2} \ . 
\end{eqnarray}


\end{document}